\documentclass[usenatbib,fleqn]{mn2e} 
\usepackage{amsmath,amssymb}
\usepackage{tablefootnote}   
\usepackage{graphicx}         
\usepackage{gensymb}        
\usepackage{color}               
\usepackage{natbib}
\usepackage{times}
\usepackage{fixltx2e} 
\usepackage[flushleft]{threeparttable}

\usepackage{breakcites}

\usepackage[breaklinks,colorlinks,urlcolor=blue,citecolor=blue,linkcolor=blue]{hyperref}
  
\newcommand{\ramses}{{\small RAMSES}}
\newcommand{\ramsestext}{{\small RAMSES} }

\newcommand{\things}{{\small THINGS }}
\newcommand{\thingstext}{{\small THINGS}}

\newcommand{\HH}{{\rm H}$_2$ }
\newcommand{\HHH}{{\rm H}$_2$}
\newcommand{\HI}{{\rm H}${\rm\scriptstyle I}$}

\newcommand{\Msol}{\,{\rm M}_\odot} 
\newcommand{\Msolyr}{{\,M}_\odot\,{\rm yr}^{-1}} 
\newcommand{\Mpc} {{\,\rm Mpc}}
\newcommand{\kpc} {{\,\rm kpc}}
\newcommand{\pc} {{\,\rm pc}} 
\newcommand{\K} {{\,\rm K}} 
\newcommand{\cc}{{\,\rm {cm^{-3}}}}

\newcommand{\kmsec}{{\,\rm {km\,s^{-1}} }}

\def\Myr{\,{\rm Myr}}
\newcommand{\bvect}[1]{\boldsymbol{#1}}

\newcommand{\Pk}{\langle P(k) \rangle}

\title[The Impact of Feedback on the ISM] {The impact of stellar feedback on the density and velocity structure of the interstellar medium}\author[Kearn Grisdale
  et al.] {\parbox[t]{\textwidth}{Kearn Grisdale$^1$\thanks{k.grisdale@surrey.ac.uk}, Oscar Agertz$^{1,2}$, Alessandro B. Romeo$^{3}$, Florent Renaud$^{1}$ and Justin I. Read$^{1}$}\vspace*{6pt}\\
  $^1$ Department of Physics, University of Surrey, Guildford GU2 7XH, United Kingdom \\ 
  $^2$   Lund Observatory, Department of Astronomy and Theoretical Physics, Box 43, 221 00 Lund, Sweden\\
  $^3$ Department of Earth and Space Sciences, Chalmers University of Technology, SE-41296 Gothenburg, Sweden\\
  }
\date{\today}
\begin{document}
\maketitle

%-------------------------------------------------------------------------------------------------------------------------------------------------------------------------------------------
%--------------------------------------------------------------------------- Section: Abstract ----------------------------------------------------------------------------------------

\begin{abstract}
We study the impact of stellar feedback in shaping the density and velocity structure of neutral hydrogen (\HI) in disc galaxies. For our analysis, we carry out $\sim 4.6$\,pc resolution $N$-body+adaptive mesh refinement (AMR) hydrodynamic simulations of isolated galaxies, set up to mimic a Milky Way (MW), and a Large and Small Magellanic Cloud (LMC, SMC). We quantify the density and velocity structure of the interstellar medium using power spectra and compare the simulated galaxies to observed \HI{} in local spiral galaxies from \things (The HI Nearby Galaxy Survey). Our models with stellar feedback give an excellent match to the observed \things \HI{} density power spectra. We find that kinetic energy power spectra in feedback regulated galaxies, regardless of galaxy mass and size, show scalings in excellent agreement with super-sonic turbulence ($E(k)\propto k^{-2})$ on scales below the thickness of the \HI{} layer. We show that feedback influences the gas density field, and drives gas turbulence, up to large (kpc) scales. This is in stark contrast to density fields generated by large scale gravity-only driven turbulence. We conclude that the neutral gas content of galaxies carries signatures of stellar feedback on all scales.
\end{abstract}


%-------------------------------------------------------------------------------------------------------------------------------------------------------------------------------------------
%--------------------------------------------------------------------------- Section: Introduction -----------------------------------------------------------------------------------
\section{Introduction}
\label{sect:intro}
Galaxy formation is an inefficient process. Different methods such as abundance matching \citep{Tasitsiomi04,vale_ostriker04,Read:2005aa,kravtsov_etal14}, satellite kinematics \citep{klypin_prada09,more_etal11}, and weak lensing \citep{mandelbaum_etal06} all indicate that a surprisingly low fraction of cosmic baryons tend up as stars in the centers of dark matter haloes, with stellar to dark matter mass fractions of $M_\star/M_{\rm h}\approx 3-5\,\%$  for $L_\star$ galaxies, well below the cosmological baryon fraction of $\Omega_{\rm b}/\Omega_{\rm m}\approx 16\%$ \citep{Planck2015}. For both less and more massive galaxies $M_\star/M_{\rm h}$ is even lower.

Such low baryon fractions are believed to be due to galactic winds driven by stellar feedback (e.g. radiative feedback, supernovae, stellar winds, cosmic rays etc.) at the faint end of the stellar mass function \citep{DekelSilk86,Efstathiou00} and by the active galactic nuclei (AGN) and the bright end \citep{SilkRees1998,Benson2003}. Over the last two decades there has been a significant effort to models these processes in simulations of galaxy formation and evolution \citep[e.g.,][]{Katz92,NavarroWhite93,ThackerCouchman2001,Stinson06,AvilaReese2011,Hopkins2014,Kimm2015,AgertzKravtsov2015}. However, stellar feedback has proven to be challenging to simulate and results are generally mixed, as many processes are still poorly understood, e.g. the efficacy of radiative feedback for galactic wind driving. No consensus yet exists on exactly which feedback processes are crucial to model. At a spatial resolution of $\sim 500\pc$, \cite[][]{Aquila} show that, for Milky Way-mass galaxies, the details of the algorithmic implementations of stellar feedback significantly influence the resulting galaxies. However, higher resolution simulations that reach spatial scales of $\lesssim100$\,pc and begin to resolve the multiphase ISM are in much better agreement \cite[e.g,][]{Kravtsov2003,Stinson2013,Hopkins2014,AgertzKravtsov2016,Read:2016ac,Read:2016aa}. These simulations find that stellar feedback allows galaxy formation simulations to reproduce not only the stellar mass-halo mass relation, but also the evolution of basic properties of spiral galaxies: stellar mass, disc size, morphology dominated by a kinematically cold disc, stellar and gas surface density profiles, and specific angular momentum.

Given the improvement in spatial resolution, with many authors now reporting sub-100 pc resolution in a cosmological context\citep[e.g,][]{Hopkins2014,AgertzKravtsov2015,AgertzKravtsov2016}, and even higher resolution for isolated galaxy simulations \citep[e.g.][]{Renaud:2013aa, Read:2016aa} this opens up the possibility of studying the role of feedback by not only considering integral properties of galaxies such as size and mass, but by using detailed observations of the structure of neutral gas in nearby galaxies \citep[e.g.][]{Walker:2014aa}.

Today, decades after the pioneering work by \cite{larson81}, observations and simulations of the interstellar medium (ISM) are revealing its turbulent nature with higher and higher fidelity \citep[see review by e.g.][]{maclow:review04,elmegreen04}. HI emission lines in most spiral galaxies have characteristic velocity dispersions of $\sigma\sim 10 {\rm\, km \,s^{-1}}$ on a scale of a few hundred parsecs \citep[][]{Tamburro2009}, exceeding the values expected from purely thermal effects. This suggests that the ISM is \emph{super-sonically} turbulent. Turbulence controls the overall structure of the ISM, and is an important ingredient for star formation \citep[][]{mckeeostriker07}, not only for determining the rate of star formation in molecular clouds \citep[][]{Federrath2012,Padoanreview2013}, but also by affecting the global and local stability properties of galaxies \citep[][]{Romeo2010, Hoffmann:2012aa,Romeo2014,Agertz:2015aa}.

The structure of the ISM has been studied using numerical simulations of isolated models of galaxies \citep[e.g.][]{taskerbryan06,Agertz09,Dobbs2011,Hopkins2012structure,Renaud:2013aa,Marasco:2015aa}, as well as individual patches of the ISM \cite[][]{Joung:2006aa,Walch2015,Martizzi:2015aa,Gatto:2015aa}. The main sources of turbulence driving are still not clear. Several candidates capable of driving the ISM turbulence exist, for example large-scale expanding outflows from high-pressure HII regions \citep{kessel03}, stellar winds or supernovae \citep[e.g.][]{kim01,deavillez04,deavillez05,Joung:2006aa}, gravitational instabilities coupled with galactic rotation \citep[][]{gammie91,piontekostriker04,Ibanez-Mejia:2015aa,Krumholz:2016aa} and the Magneto-Rotational-Instability (MRI) \citep{balbushawley91}, to name a few.

The turbulent ISM is often conveniently characterised in Fourier space using power spectra. \cite{Kolmogorov:1941aa} proved that incompressible turbulence will feature an energy spectra $E(k)\propto k^{-5/3}$ below the injection scale, where $k\propto 1/\ell$ is the wavenumber and $\ell$ the physical scale. Supersonically turbulent gas features a steeper spectra, with $E(k)\propto k^{-2}$ \citep{Burgers:1948aa}, as conformed by idealised numerical simulations of turbulent gas \cite[e.g.][]{Kritsuk:2007aa}. On galactic scales, simulations indicate that turbulent scalings are present \citep[e.g.][]{wada02,Agertz:2015aa}, but as mentioned above, {\bf it is} not established which physical mechanisms are required to maintain them, and whether the observed \HI{} content of galaxies feature the same energetics as idealised galaxy simulations. Observations of the neutral ISM in nearby galaxies have revealed that the \HI{} component often gives rise to simple power law behaviours ($P(k)\propto k^{-\alpha}$), with $\alpha$ typically in the range $1-3$, over several ordered of magnitude in scale \citep[e.g.][]{Stanimirovic:1999aa,Bournaud:2010aa, Combes:2012aa,Zhang:2012aa,Dutta:2013aa}. This points towards the presence of turbulence. Broken power laws are also observed, possibly owing to the finite thickness of the disc causing a transition from 2D turbulence on large scales to 3D turbulence on small scales \citep[for example, see NGC 1058 in][]{Dutta:2009aa}.

This leads us to the main question that we seek to answer in this work: \emph{what are the observational signatures of stellar feedback on the galactic density and velocity structure of \HI{} gas, from small ($\sim 10\pc$) to large ($>\kpc$) scales, and can these be used to discriminate between sources of turbulence?} To address this question, we perform hydrodynamical simulations of disc galaxies of varying sizes and masses (thus giving different star formation rates and therefore different feedback injection rates) and compare these to observations of the cold ISM in nearby galaxies. Our goal is to understand the role of stellar feedback in shaping the density and velocity structure of the ISM. 

The paper is organised as follows. In \S\ref{sect:meth} we describe the numerical method, the observational data, as well as the adopted power spectrum analysis method. In \S\ref{sect:results} we present power spectra of the simulated density and energy field and compare these to the observed neutral ISM in nearby spiral galaxies. We discuss our findings in \S\ref{sec:discussion} and investigate existing uncertainties in simulations and observations. Furthermore, we relate our results to the literature. Finally, in \S\ref{sect:con} we present our conclusions.

\begin{table*}
	\begin{center}
		\parbox{0.93\textwidth}{	
			\caption{Initial conditions of simulated galaxies}
			\begin{tabular}[h]{c r r r c c c c c c c c }
				\hline \hline
				Simulation & Stellar disc & Stellar bulge & Dark matter  & Gas        & Scale   & Scale & Bulge scale & $v_{\rm 200}$  & $c$ & Gas metallicity\\
				                  & mass          & mass            & halo mass     & fraction  & length  & height & length          &                         &        &                 \\
				                  & [$10^8$M$_{\odot}$] &  [$10^8$M$_{\odot}$] & [$10^{10}$M$_{\odot}$] &  & [kpc] & [kpc] & [kpc] & [km s$^{-1}$]  &  &  [$Z_\odot$]           \\
				\hline
				MW       & 343.7  &  43.0      & 125.4   & 0.20   &  3.43 & 0.34  & 0.3432 &  150  &  10  & 1.75\\
				LMC      &   21.8  &    0.035  &   21.28 & 0.38   &  3.35 & 0.34  & 0.0167 &    82  &   9   & 0.53\\
				SMC      &    1.1   &    0.005  &    2.59  & 0.90  &  1.16  & 0.12  & 0.0058 &    42  & 15   & 0.18\\ 
				\hline 
				\hline

			\end{tabular}
			{\footnotesize
			Notes: {\bf Column 1:} name of simulation, {\bf Column 2:} stellar mass of  the disc, {\bf Column 3:} stellar mass of the galactic bulge, {\bf Column 4:} mass of dark matter halo, {\bf Column 5:} gas to stellar mass ($M_{\rm gas,disc}/M_{\rm \star,total}$), {\bf Column 6:} gas scale length of the disc, {\bf Column 7:} gas scale height of disc, {\bf Column 8:} gas scale radius of galactic bugle, {\bf Column 9:} rotational velocity at r$_{200}$, {\bf Column 10:} concentration parameter and {\bf Column 11:} gas metallicity as a function of solar metallicity.
			}
			\label{tab:IC}
		}
	\end{center}	
\end{table*}

%-------------------------------------------------------------------------------------------------------------------------------------------------------------------------------------------
%--------------------------------------------------------------------------- Section: Method -----------------------------------------------------------------------------------------
\section{Method}
\label{sect:meth}
%--------------------------------------------------------------------------- Section: Method: Numerical Technique -----------------------------------------------------------
\subsection{Numerical technique}
\label{sec:meth_sims}
We use the Adaptive Mesh Refinement (AMR) code {\ramsestext} \citep{Teyssier:2002aa} to carry out hydro+$N$-body simulations of galactic discs for comparison with observations. The fluid dynamics of the baryons is calculated using a second-order unsplit Godunov method, while the collisionless dynamics of stellar and dark matter particles is evolved using the particle-mesh technique \citep{Hockney1981}, with gravitational accelerations computed from the gravitational potential on the mesh. The gravitational potential is calculated by solving the Poisson equation using the multi-grid method \citep{GuilletTeyssier2011} for all refinement levels. The equation of state of the fluid is that of an ideal mono-atomic gas with an adiabatic index $\gamma=5/3$. 

The code achieves high resolution in high density regions using adaptive mesh refinement, where the refinement strategy is based on a quasi-Lagrangian approach in which the number of collisionless particles per cell is kept approximately constant at 8. This allows the local force softening to closely match the local mean inter-particle separation, which suppress discreteness effects \citep[e.g.,][]{Romeo08}. In addition to this, cell refinement is triggered when the baryonic mass (stars and gas) in a cell exceeds a fixed threshold $M_{\rm ref}$, see \S\ref{sect:simsuite}.

The star formation, cooling physics and stellar feedback model adopted in our simulations is identical to the implementation used in \cite{Agertz:2015aa}, and described in detail in \cite{Agertz:2013aa} and \cite{AgertzKravtsov2015} \citep[see also][]{Read:2016ab,Read:2016aa} . We refer the reader to those papers for details. Briefly, we model the local star formation rate using the following equation,
\begin{equation}
	\label{eq:schmidtH2}
	\dot{\rho}_{\star}=f_{\rm H_2}\frac{ \rho_{\rm g}}{t_{\rm SF}},
\end{equation}
where $f_{\rm H_2}$ is the local mass fraction of molecular hydrogen (H$_2$), $\rho_{\rm g}$ is the gas density in a cell, and $t_{\rm SF}$ is the star formation time scale of {\it molecular} gas. The fraction of molecular hydrogen in a cell is a function of the gas density and metallicity and is computed using the KMT09 model \citep[][]{kmt08,KMT09} which implemented as described in \cite{AgertzKravtsov2015} (see their \S2.3, equations 2-6). 

The star formation time scale is related to the {\it local} efficiency of star formation in a computational cell of a given density as $t_{\rm SF}=t_{\rm ff, SF}/\epsilon_{\rm ff, SF}$, where $t_{\rm ff, SF}=\sqrt{3\pi/32G\rho_{\rm g}}$ is the local free-fall time of the star forming gas and $\epsilon_{\rm ff, SF}$ is the local star formation efficiency per free-fall time. We adopt $\epsilon_{\rm ff, SF}=1\%-10\%$ in this work, see \S\,\ref{sect:simsuite}. In every cell, Eq.\,\ref{eq:schmidtH2} is sampled using a Poisson process \citep[see e.g.][]{dubois08}, where resulting star particles are assumed to have an internal stellar mass distribution according to a \cite{Chabrier:2003aa} initial mass function (IMF). At the time of formation star particles  have an initial mass of $300\Msol$, and can lose up to $\sim 40\%$ of their mass over a Hubble time due to stellar evolution processes.

Several processes contribute to the stellar feedback budget, as stars inject energy, momentum, mass and heavy elements over time via SNII and SNIa explosions, stellar winds and radiation pressure into the surrounding gas. Metals injected by supernovae and stellar winds are advected as a passive scalar and are incorporated self-consistently in the cooling and heating routine. Furthermore, we adopt the SN momentum injection model recently suggested by \cite{Kim:2015aa} \citep[see also][]{Martizzi:2015aa,Gatto:2015aa,Simpson:2015aa}. 
Here we consider a SN explosion to be resolved when the cooling radius\footnote{the cooling radius scales as $r_{\rm cool}\approx 30 n_0^{-0.43}(Z/Z_\odot+0.01)^{-0.18}$ pc for a supernova explosion with energy $E_{\rm SN}=10^{51}$ erg \citep[e.g.][]{Cioffi1988,Thornton1998,Kim:2015aa}} is resolved by at least three grid cells ($r_{\rm cool}\geq3\Delta x$). In this case the explosion is initialised in the energy conserving phase by injecting the relevant energy ($10^{51}\,$erg per SN) into the nearest grid cell. If this criterion is not fulfilled, the SN is initialised in its momentum conserving phase, i.e. the total momentum generated during the energy conserving Sedov-Taylor phase is injected into to the 26 cells surrounding a star particle. It can be shown \citep[e.g.][]{Blondin1998,Kim:2015aa} that at this time, the momentum of the expanding shell is approximately $p_{\rm ST}\approx 2.6\times 10^5\,E_{51}^{16/17}n_0^{-2/17} \Msol\kmsec$.

%------------------------------- Table: THINGS comparisons --------------------------------------------
\begin{table*}
	\begin{center}
		\parbox{1.06\textwidth}{
			\caption{Summary of our analysed sample of galaxies from \things}
			\begin{tabular}[h]{c c c c c c c c l l l }		
				\hline \hline
				Galaxy & Distance & Inclination & $M_{\rm HI}$  &  $M_{\rm H_2}$  &  $M_{\star}$  & Global & \HI{} beam & Comments \\
				&&&&&& SFR & width \\
	  			& [Mpc] & [\degree] & [$10^8M_{\odot}$] & [$10^8M_{\odot}$] & [$10^8M_{\odot}$] & [M$_\odot$ yr$^{-1}$] & [pc] \\
				\hline
				NGC 628   &   7.3 &   7 &   38.0 &    8.3 &  37.15 & 1.21  & 240 & LMC-like; face-on; medium \HI-fraction and low $M_\star$\\
				NGC 3521 & 10.7 & 73 &   80.2 &  26.5 & 602.56 & 3.34 & 425 & MW-like; edge-on; high \HI-fraction and $M_\star$\\
				NGC 4736 &   4.7 & 41 &     4.0 &    4.1 & 218.78 & 0.43 & 136 & MW-like; face-on; low \HI-fraction and medium $M_\star$\\
				NGC 5055 & 10.1 & 59 &   91.0 &  36.2 & 575.44 & 2.42 & 263 & MW-like; edge-on; high \HI-fraction and $M_\star$\\
				NGC 5457 &   7.4 & 18 & 141.7 &  19.8 & 107.15 & 1.49 & 269 & MW-like; face-on; low \HI-fraction and $M_\star$\\
				NGC 6946 &   5.9 & 33 &   41.5 &  32.0 &   91.20 & 4.76 & 173 & MW-like; face-on;  medium \HI-fraction and low $M_\star$\\
				\hline
				\hline
			\end{tabular}
			\label{table:things}\\
			[\smallskipamount]
			{\footnotesize
			Notes: {\bf Column 1:} names of galaxy, {\bf Column 2:} Distance to galaxy \citep[see][]{THINGSpaper}, {\bf Column 3:} Inclination of galaxy (NB: $0\degree=$ face-on) \citep[see][]{THINGSpaper}, {\bf Column 4:} \HI{} mass \citep[see][]{THINGSpaper}, {\bf Column 5:} \HH mass (calculated from column 3, Table 5 of \cite{THINGSpaper} and Table 3 column 7 of \cite{Leroy:2009aa}), {\bf Column 6:} stellar mass \citep[see Table 1, column 8][]{Skibba:2011aa}, {\bf Column 7: } global star formation rate \citep[see Table 1, column 10 of][]{Skibba:2011aa}, {\bf Column 8:} \HI{} beam width \citep[calculated from][]{THINGSpaper}, {\bf Column 9:} simulation the observed galaxy best matches; viewing perspective; \HI{} mass and stellar mass compared to the MW simulation. 
			}
		}
	\end{center}	

\end{table*}
%------------------------------------------------------------------------------------------------------------------

%--------------------------------------------------------------------------- Section: Method: Simulation Suite  -----------------------------------------------------------------
\subsection{Simulation suite}
\label{sect:simsuite}
We run reference simulations, in which density and turbulence are set only by gravity and hydrodynamics (i.e. with no feedback regulation). We then quantify the effect of stellar feedback in shaping the ISM by running identical simulations including the stellar feedback model described above and compare the density and turbulence structures of the two sets of simulations.

We carry out numerical simulations of Milky Way (MW), Large Magellanic Cloud (LMC) and Small Magellanic Cloud (SMC)-like galaxies. The characteristics of these galaxies are presented in Table~\ref{tab:IC}. The initial conditions (ICs) feature a stellar disc, stellar bulge, gaseous disc and dark matter halo. We set up the particle distributions following the approach by \cite{Hernquist:1993aa} and \cite{Springel:2000aa} \citep[see also][]{Springel:2005aa}, assuming an exponential surface density profile for the disc, a Hernquist bulge density profile \citep{Hernquist1990}, and an NFW dark matter halo profile \citep{nfw1996}. We use $10^6$ particles for both the NFW halo and stellar discs, with the same mass resolution in the bulge component as in the disc. We initialise the gaseous disc on the AMR grid assuming an exponential profile, and assume the galaxies to be embedded in a hot ($T=10^6\K$), tenuous ($n=10^{-5}\cc$) corona enriched to $Z=10^{-2}Z_\odot$, while the discs have the  abundances given in Table \ref{tab:IC}. All simulations include the same tenuous hot corona. Despite the temperature and density being reasonable for the Milky Way \citep[see ][]{Gatto:2013aa}, the corona in all simulations is unrealistic, for example the gas is not stratified. However, as the coronal mass is insignificant, and there is no significant accretion onto the discs, these corona models are sufficient for this work. Each galaxy is simulated in isolation, i.e. we neglect environmental effects such as galaxy interactions. The galaxies are set at the centre of a box with a size of $L_{\rm box}=600\kpc$, and run with 17 levels of adaptive mesh refinement, allowing for a finest grid cell size of $\Delta x\sim 4.6$pc.  The mass refinement threshold is $M_{\rm ref}\approx 9300 \Msol$ for the MW simulations and $M_{\rm ref}\approx 930 \Msol$ for LMC and SMC simulations, leading to baryon mass resolutions of just below $\sim 1200\Msol$ and $\sim 120\Msol$  respectively.

The MW models use the ICs from \cite{Agertz:2013aa} \citep[the AGORA ICs, see also][]{agora}. We generate the LMC and SMC ICs by applying the parametrisation adopted by \cite{Mo:1998aa} using the following parameters from Table 1 of \cite{Besla:2010aa}: concentration, gas fraction and $V_{200}$ (the virial velocity, i.e. the circular velocity at the radius where the mean density of the dark matter halo is 200 times the critical density of the Universe). For both the LMC and SMC simulations we employ a spin parameter $\lambda=0.05$ \citep[e.g. ][]{Bullock:2001aa}. We emphasise that these simulations are not designed to match the real galaxies, instead we model spiral galaxies with similar gas fractions, stellar masses and dark matter halo masses as the Milky Way, LMC and SMC. While the ICs employed for the LMC and SMC-like simulations have stellar and gas masses consistent with observations of the LMC \citep[][]{Meatheringham:1988aa,Kim:1998aa,Marel:2002aa} and SMC \citep[][]{Stanimirovic:2004aa,Yozin:2014aa}, we note that the adopted scale lengths for the initial gas distributions are smaller than what is commonly derived from observations \cite[see e.g.][and references within]{Besla:2010aa}.

We note that the MW analogue was designed to have characteristics of a typical Sb-Sbc galaxy in order to facilitate a comparison with the \HI{} data in the \things spiral galaxy sample, see \S\ref{sect:obs}. The LMC and SMC models allows us to study how the ISM is influenced by stellar feedback in low mass galaxies, and will be compared both to previous numerical studies \citep[e.g.][]{Bournaud:2010aa} and observations \citep[e.g.][]{Stanimirovic1999}.

The gas discs in our simulated galaxies feature a rather  elevated oxygen abundance compared with observed values, leading to the average metallicities shown in Table \ref{tab:IC}. In the case of the MW model, the adopted metallicity is more representative of the inner disc ($R<R_\odot$) rather than at the solar radius. We have confirmed that this does not affect the conclusions of this paper by re-simulating our LMC-like galaxy with a lower metallicity ($Z =0.3\,Z_\odot$).

For the no feedback models (denoted `noFB') we adopt a local star formation efficiency per free-fall time of $\epsilon_{\rm ff}=1\%$. This low efficiency, motivated by the results of e.g. \cite{krumholztan07}, leads to a galaxy matching the empirical $\Sigma_{\rm SFR}-\Sigma_{\rm gas}$ (Kennicutt-Schmidt, KS) relation \citep[][]{Kennicutt:1998aa,Bigiel:2008aa}, as shown by \cite{Agertz:2013aa}, and implicitly assumes regulated star formation, albeit without the explicit action of stellar feedback. In contrast, in the stellar feedback regulated galaxy models (denoted `FB') we adopt a larger efficiency,  $\epsilon_{\rm ff}=10\%$, allowing for feedback to regulate the star formation process back to the observed low efficiencies \citep[e.g.][]{AgertzKravtsov2016}, i.e. reproducing the empirical KS relation, while shaping the ISM in the process. These two different models of galaxy evolution (with and without stellar feedback) allow us to investigate the role of feedback in shaping the ISM.

\subsection{Observational data}
\label{sect:obs}

\label{sect:meth_things}
We make use of \HI{} data from The \HI{} Nearby Galaxy Survey (\thingstext) \cite{THINGSpaper}. The survey focused on galaxies within a distance of $15\Mpc$ at a resolution of $\sim 6^{\prime\prime}$, resulting in a spatial resolution (beam size) of $100\lesssim \ell \lesssim 500$ pc. We select six spiral galaxies from \thingstext, all observed close to face-on except for NGC 3521 and NGC 5055\footnote {For our analysis we treat \emph{all} \things data as face-on galaxies, but explore the role of inclination on our analysis with simulations in \S\ref{sect:den_inc}, Fig.~ \ref{fig:All_Obs_sim_PS_comp}, \S\ref{sect:vel_smooth} and Fig.~\ref{fig:KE_things_sim_comp}}. Despite having larger inclination angles, both \HI{} surface density maps of NGC 3521 and NGC 5055  still show clear spiral structure. These six galaxies where selected to give a range in the \HI{} gas mass and similar SFR to our simulated MW-like galaxy. In Table \ref{table:things} we present the distances, inclination angles, resolution limits, star formation rates (SFRs), and \HI , \HH and stellar masses for our sample.

We use the `robust weighting' data maps from \things as these offer higher resolution and a more uniform beam size, see \cite{THINGSpaper} for a detailed discussion. The zero moment maps are used for calculating properties of the \HI{} density field, such as the two dimensional density power spectra (see \S\,\ref{sect:meth_PS}), and the first moment maps for our analysis of the line of sight \HI{} velocity field.

%--------------------------------------------------------------------------- Section: Method: Computing Power Spectra ----------------------------------------------------
\subsection{Computing power spectra}
\label{sect:meth_PS}

Quantifying the density structure and kinetic energy of the cold ISM using power spectra is common practise in astrophysics, often in the context of turbulent flows, see discussion in \S\ref{sect:intro}. Throughout this work we present density and energy power spectra analysis from both simulations and observations, where we have opted to analyse both sources of data in a similar way as possible to allowing for a direct comparison. As we are working with discrete data we use the  `Fastest Fourier Transform in the West' ({\small FFTW}\footnote{\url{http://www.fftw.org}}) routines \citep{FFTW} to calculate the discrete Fourier transform. The power spectrum is defined as
\begin{equation}
	P(\bvect{k}) = \tilde{\bvect{w}}(\bvect{k}) \cdot \tilde{\bvect{w}}(\bvect{k})^*
	\label{eq:pk}
\end{equation}
where $\tilde{\bvect{w}}(\bvect{k})$ is the Fourier transform of the real array $\bvect{w}(\bvect{x})$ and $\bvect{k}$ is the wave vector. Here `*' refers to the complex-conjugate. To obtain an isotropic (one dimensional) power spectrum we bin $P(\bvect{k})$ in wave vectors $k=\left|\bvect{k}\right|$ and divide each bin by the number of contributing data points ($N_k$).  This gives $\langle P(k) \rangle$ which is commonly referred to as the ``angle-averaged power spectrum'' \citep[e.g.][]{Joung:2006aa}. The wave vector $k = 2\pi / \ell$, where $\ell$ is a physical scale. This allows us to define an \emph{energy spectrum} of the quantify $\bvect{w}(\bvect{x})$ as
\begin{equation}
	\label{eq:1dspectrum}	
	E(k) \equiv \pi (2 k)^{(D-1)} \langle P(k)\rangle,
\end{equation}
where $D$ is the number of dimensions of the input data. 

\begin{figure*}
		\includegraphics[width=0.9\textwidth]{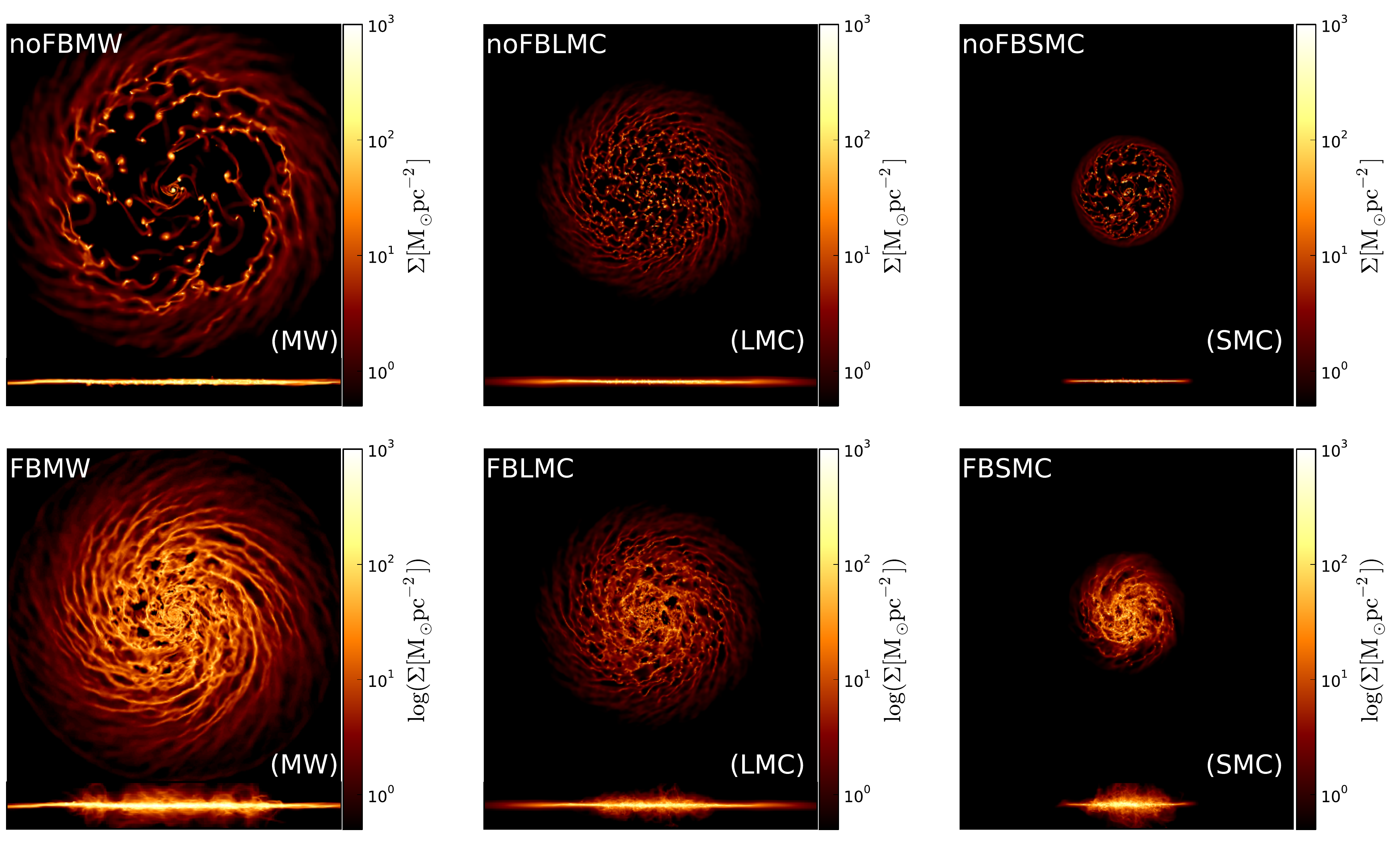}
		\caption{Surface density maps of our simulations at $t=450$Myr. The top row shows simulations without feedback while the bottom shows those with feedback. From left to to right we show our MW, LMC and SMC-like simulations. Each panel shows the face-on view ($36\times36\kpc$) above the edge-on view ($36\times5$kpc) the simulation. 		}
		\label{fig:Maps}
\end{figure*}

%--------------------------------------------------------------------------- Section: Method: Surface Density PS -------------------------------------------------------------
\subsubsection{Surface density power spectra}
\label{sect:meth_dpow}

We quantify the density structure of the observed and simulated ISM by computing the power spectra of \HI{} surface densities following the definition of $\Pk$ above. 

For the observations we convert the \HI{} data into units of $\Msol\pc^{-2}$ using the `robust weighted density map' (see \S\ref{sect:obs}) using equation 3 from \cite{THINGSpaper}, combined with the small angle approximation, ignoring inclination effects (we explore the effect of inclination on our results using the simulation data in \S\ref{sect:den_inc}, Fig.~ \ref{fig:All_Obs_sim_PS_comp}, \S\ref{sect:vel_smooth} and Fig.~\ref{fig:KE_things_sim_comp}). We use the entire data map, i.e. we do not limit our analysis to a subregion of the maps, therefore our analysis of observational data takes the entire \HI{} disc into account.

For the simulations we calculate surface density maps assuming the galaxies are 'observed' face-on, i.e. along the axis of rotation, but explore inclination effects in the  \S\ref{sect:den_inc}, Fig.~ \ref{fig:All_Obs_sim_PS_comp}, \S\ref{sect:vel_smooth} and Fig.~\ref{fig:KE_things_sim_comp}. We compute the maps for the simulations at a uniform resolution of $\Delta x=4.6\pc$, unless otherwise stated. Each map has size of $36\times 36\kpc^2$, centred on the galaxy centre. The \HI{} map is generated by computing the \HI{} fraction in each cell assuming collisional ionisation equilibrium (CIE). 

To avoid contamination arising from the inherent periodic boundary conditions from the FFT technique, we pad both the simulation and observational data by placing the maps in the centre of a square void domain (zero-padding). The size of the void domain in each dimension is set to $2^n$ where $n$ is the first integer where $2^n$ is greater than the number of cells of the surface density map in the same dimension. Experiments with more padding yielded identical results.

\begin{figure*}
	\begin{center}
	\begin{tabular}[h]{ll}
		\includegraphics[width=0.45\textwidth]{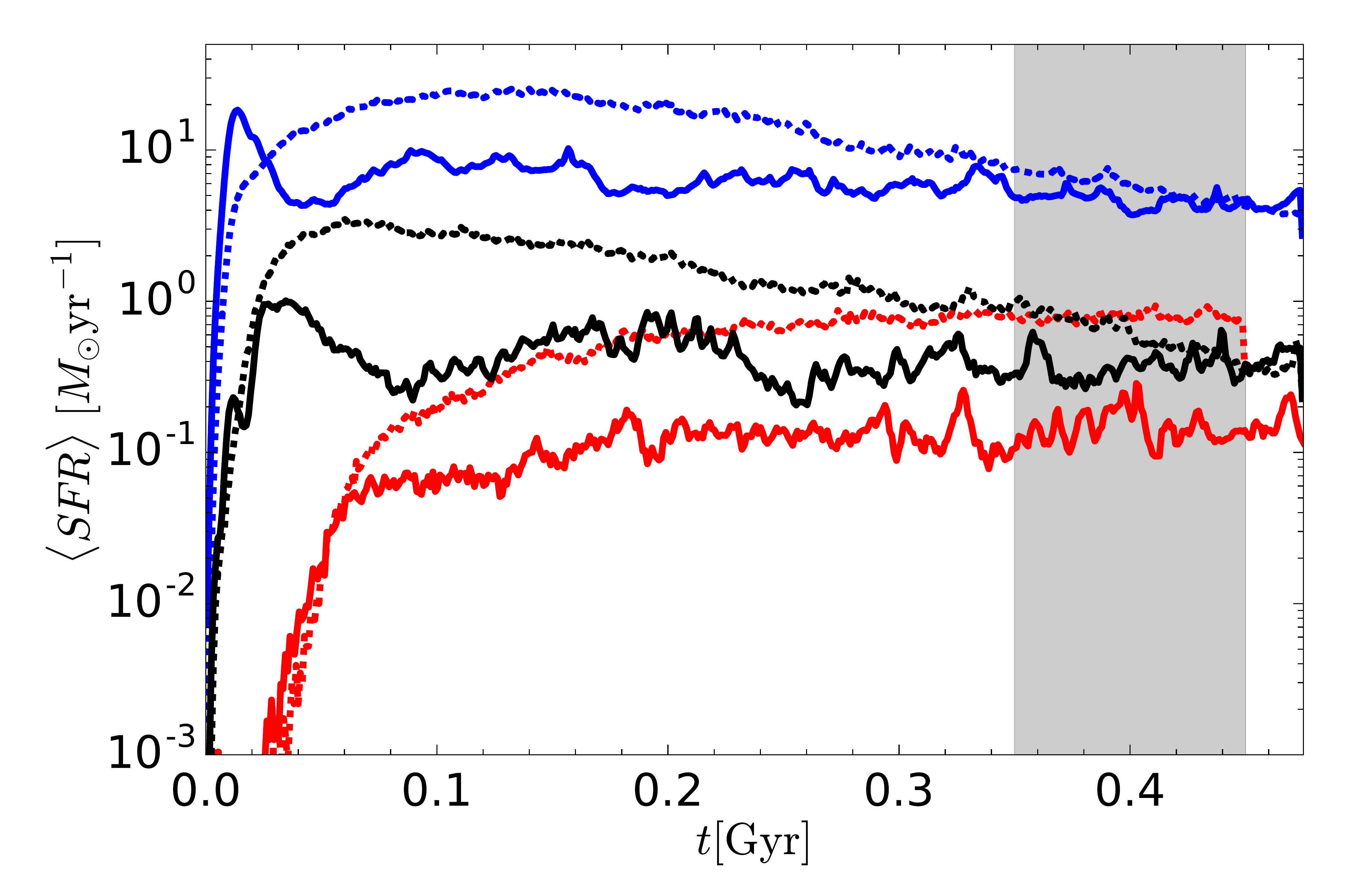}
		\includegraphics[width=0.45\textwidth]{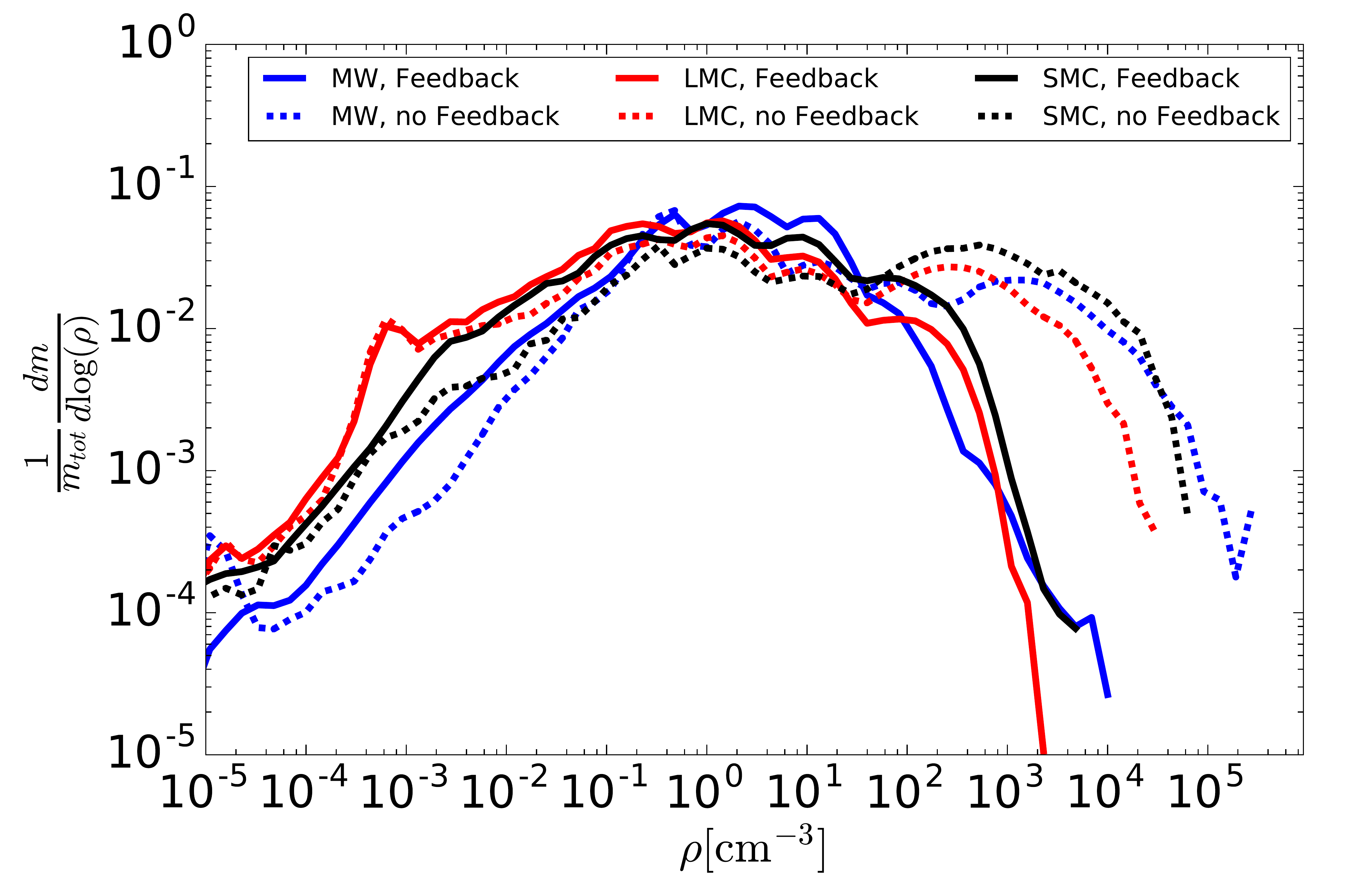}	
			\end{tabular}
			\caption{Mean global SFR (left) and Probability Density Function (PDF) (right) for all simulations at $t=350$Myr. Feedback runs are shown with solid lines, while no feedback runs are shown with a dashed line. Blue, red and black represent our Milky Way, LMC and SMC simulations respectively. The shaded region on the left panel shows the period of our analysis, see \S\ref{sect:SDPS}
		}
		\label{fig:SFHandPDF}
	\end{center}
\end{figure*}

%--------------------------------------------------------------------------- Section: Method: Kinetic Energy PS ---------------------------------------------------------------
\subsubsection{Kinetic energy power spectra}
\label{sect:meth_vpow}
We study the energetics of the ISM by computing 2D kinetic energy power spectra ($E_{\rm KE,2D}(k)$) from the simulations and observations, as well as 3D spectra ($E_{\rm KE,3D}(k)$) for the simulations. In both the 2D and 3D case we adopt the definition of the energy spectrum $E(k)$ in equation \ref{eq:1dspectrum}.

For the 3D spectra, we account for compressibility following \cite{Kritsuk:2007ab}. i.e. the variable entering the power spectrum calculations is $\bvect{w}=\rho^{1/2}\bvect{v}$, rather than simply the velocity, where $\bvect{v}$ is the 3D \emph{turbulent} velocity vector in each simulation cell, computed by removing the average galactic rotation at the cell radius, and $\rho$ is the total gas density. We adopt this definition, as this gives us the Fourier transform of the kinetic energy field, but we could instead have used $\bvect{w}=\rho^{1/3}\bvect{v}$, often described as the kinetic energy flux, which has been shown to reproduce energy spectra with Kolmogorov scaling ($E(k)\propto k^{-5/3}$) in super-sonic flows  \citep{Kritsuk:2007ab}.

For computational feasibility, we constrain our analysis to a $10\times10\times5\kpc^3$ region\footnote{Experiments at lower resolution ($\Delta x\sim18.3$) with a $10\times10\times10\kpc^3$ region produce a spectrum with almost identical shape but a slight reduction in power at all scales.} 
 in the plane of the galaxy, centred on the galactic centre. For all energy spectra we use a uniform resolution of $\sim9.2\pc$, unless otherwise stated. 
We add padding in each dimension identically to padding added to the surface density maps, then pass the region through FFTW. Due to the non-periodic nature of the disc (in the $z-$axis)\footnote{All of the 3D data of interest in the $z-$axis is contained within the cube}, we convolve the input data ($\bvect{w}$) with a Hanning window function\footnote{We use the form $h(z) = 0.5(1+\cos (2\pi z/H))$ where $h(z)$ is the window function, $z$ is the hight above the disc and $H$ is the extent of the function.} before computing the 3D energy spectrum, where the window function extends to $|z|\leq 2 \kpc$.

In 2D, relevant for a direct comparison between simulations and observations, we calculate a mass weighted line of sight velocity map\footnote{Experiments using the line-of-sight velocity dispersion of the gas instead of $v_{\rm LOS}$ yield similar result.} for the region described above (i.e. $\bvect{v}\rightarrow v_{\rm los}$). The velocity map is combined with with \HI{} surface density map to give $\bvect{w}=\Sigma_{\rm H {\tiny I}}^{1/2}v_{\rm los}$, padding is added in the each dimension of the map before from we calculate $E_{\rm KE,2D}(k)$. 

For the observed $E_{\rm KE,2D}(k)$ we use the first moment (velocity) maps for the \things data sets with bulk line of sight motion of the galaxy subtracted. As with the observed \HI{} density maps, we use the entire first moment map when calculating  $E_{\rm KE,2D}(k)$ for any of our \things galaxies.

\begin{figure*}
	\begin{center}
		\includegraphics[width=0.9\textwidth]{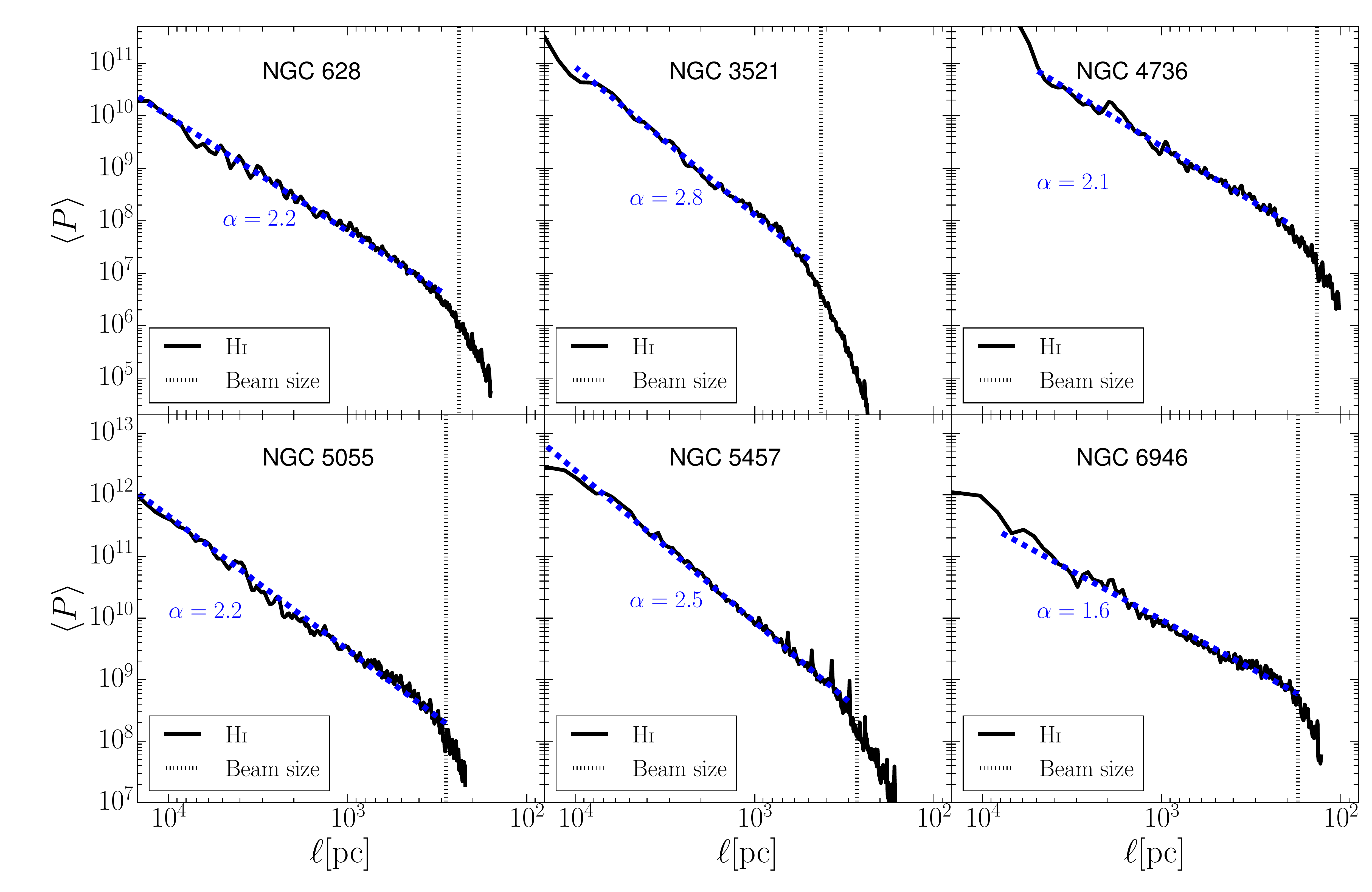}
		\caption{\HI{} surface density power spectra of the 6 galaxies from the \things sample (solid black lines). The beam size is shown by the dotted vertical lines (see Table \ref{table:things} for values). The blue lines show fitted power laws. The spectra have been normalised (see text for details)
		}
		\label{fig:DPS}
	\end{center}
\end{figure*}

%-------------------------------------------------------------------------------------------------------------------------------------------------------------------------------------------
%--------------------------------------------------------------------------- Section: Results -----------------------------------------------------------------------------------------
\section{Results}
\label{sect:results}

%--------------------------------------------------------------------------- Section: Results: Overview of Simulation results -----------------------------------------------
\subsection{Overview of simulation results}
\label{sect:generalISM}
In Fig.~\ref{fig:Maps} we show edge-on and face-on views of the projected gas density field of the simulated galaxies, with and without stellar feedback, at $t=350\Myr$. We find clear differences between each of our simulations; the feedback simulations feature galactic winds and fountains, as seen in the edge-on views, and an irregular ISM featuring feedback driven holes and transient star forming clouds. The simulations without feedback show little vertical structure as the gas cools down to thin cold discs which fragments into an ensemble of dense rapidly star forming clouds. 

In the left-hand panel of Fig.~\ref{fig:SFHandPDF} we show the resulting star formation histories of the simulated galaxies. After an initial transient all models form stars at a roughly constant rate, where the feedback regulated galaxies are less efficient in forming stars, despite a local star formation efficiency per free-fall time being 10 times greater ($\epsilon_{\rm ff}=10\%$ vs. 1\%). In subsequent sections, we carry out all of our analysis at late times ($t>350\Myr$) when the galaxies have relaxed to a new equilibrium configuration. 

Fig.~\ref{fig:SFHandPDF} shows that the SMC-like simulations have higher global SFRs than the LMC-like simulations, whereas the opposite is found in observations \citep[see][]{Kennicutt:1986aa,Wilke:2004aa,Whitney:2008aa}. The origin of this discrepancy is the adopted initial conditions. The LMC and SMC-like simulations have comparable total gas masses, but are both $\sim3\times$ more compact than the observed LMC and SMC \citep[see Table 1 of ][]{Besla:2010aa}. In addition, the SMC's smaller size leads to higher gas densities than in the LMC models, which coupled to a non-linear star formation law ($\dot{\rho}_\star\propto \rho_{\rm gas}^{1.5}$) results in higher global SFRs. 

Our MW-like simulation with feedback features a SFR of $\sim4\Msolyr$ at   $t=450\Myr$, which is a factor of a few greater than observed \citep[][found $\sim1.65\Msolyr$ for the Milky Way]{Licquia:2015aa}. In the LMC-like simulation with feedback we find a SFR $\sim0.1-0.2\Msolyr$ at $t=450\Myr$, which is a close match to values derived from observations of the LMC, $\sim0.14\Msolyr$ \citep[][]{Murray:2009aa} and $\sim0.25\Msolyr$ \citep[][]{Whitney:2008aa}, thus showing that despite the differences between our simulations and observations, at late times ($t\gtrsim 200$) our MW-like and LMC-like simulations with feedback produce global star formation rates that are compatible with those measured from observations.

For our SMC feedback simulation we find a SFR $\sim0.3-0.4\Msolyr$ at $t=450\Myr$. As noted above, these values are higher than those observed for the real SMC \citep[][found a value of $0.05\Msolyr$]{Wilke:2004aa}, by a factor of $\sim8$. We remind the reader that we do not aim to recreate an exact match to the real galaxies but instead to produce galaxies with similar gas mass.

With the exception of our LMC-like simulations, our runs without feedback reach similar SFRs to their feedback counter parts by $t=350\Myr$ i.e. at the beginning of our analysis period. The LMC-like run without feedback has a SFR $\sim4-5$ time that of its feedback counter part.

In the right-hand side of Fig.~\ref{fig:SFHandPDF} we show the Probability Distribution Functions (PDFs) for the gas density fields. For an isothermal ISM, numerical and analytical work has shown that the PDF follows a log-normal distribution \citep[e.g.][]{Vazquez:1994aa,Nordlund:1999aa,Wada:2001aa,wada07}, or a superposition of multiple log-normal distributions, each corresponding to a separate gas-phase \citep[e.g.][]{RobertsonKravtsov08}. If self-gravity of the ISM is resolved, then the PDF should develop a power-law tail at high densities \citep[in the Milky Way simulations by][ this occurs for $n\gtrsim 2000 {\rm cm}^{-3}$]{Renaud:2013aa}.

The effect of stellar feedback is striking; without feedback, gas condenses into dense clouds reaching densities of $n\sim 10^5\cc$ in all models, albeit slightly lower for the LMC model. Additionally, the PDFs for simulations without feedback do not match  log-normal distributions found in previous work of isolated galaxies, but instead resembles the PDFs of galaxies undergoing mergers \citep[see e.g.][]{Renaud:2014aa}. In contrast, in the feedback regulated simulations star forming clouds are dispersed and gas is returned to the ISM in a phase characterised by densities in the range $1\cc\lesssim n\lesssim 10\cc$. Gas here reaches maximum densities around $n\sim 10^3\cc$, two order of magnitudes less than in the models neglecting feedback, although most of the dense gas, by mass, reach only densities on the order of average GMC densities, i.e. $n\sim$ few $100\cc$.

In the next sections we will quantify how feedback affects the ISM as a function of scale and compare this to observations. 

\begin{figure*}
	\begin{center}
		\includegraphics[width=1.0\textwidth]{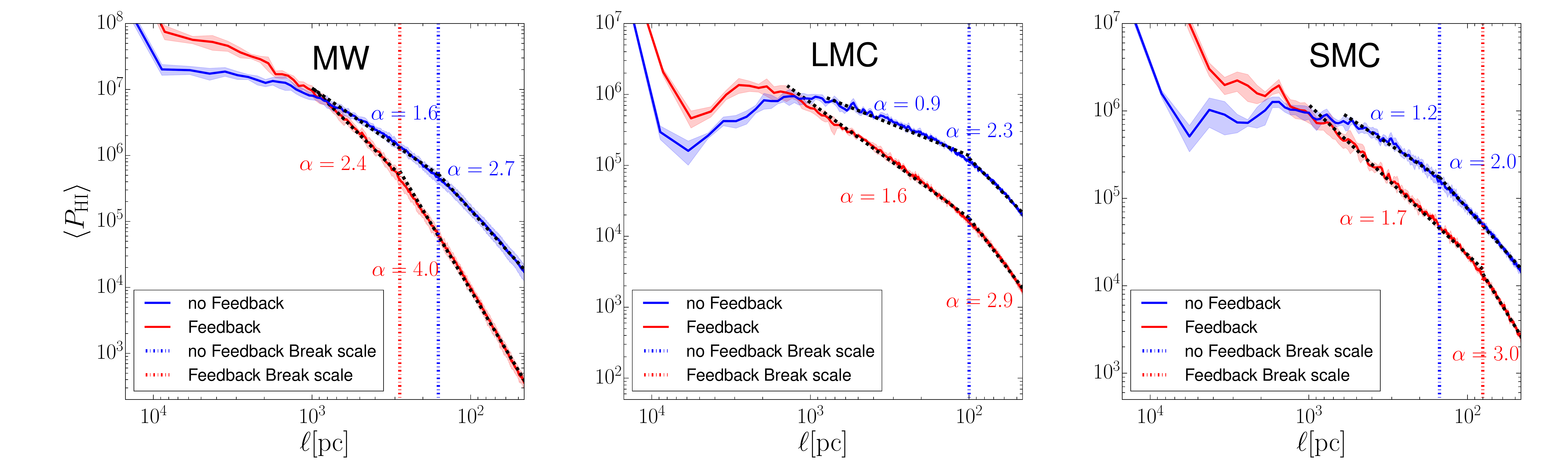}
		\caption{Time-averaged surface density power spectra using \HI{} gas for our simulations. From left to right we show the resulting spectra from the MW, LMC and SMC models with feedback shown in red and no feedback shown in blue. The dashed black lines show our power law fits with gradients next to each fit. The red and blue vertical dot-dashed lines indicate the position of the break in the power law. The shaded regions around each power spectrum show the $1\sigma$ deviation from the mean.
		}
		\label{fig:PS_Density}
	\end{center}
\end{figure*}

\begin{figure*}
		\includegraphics[width=0.9\textwidth]{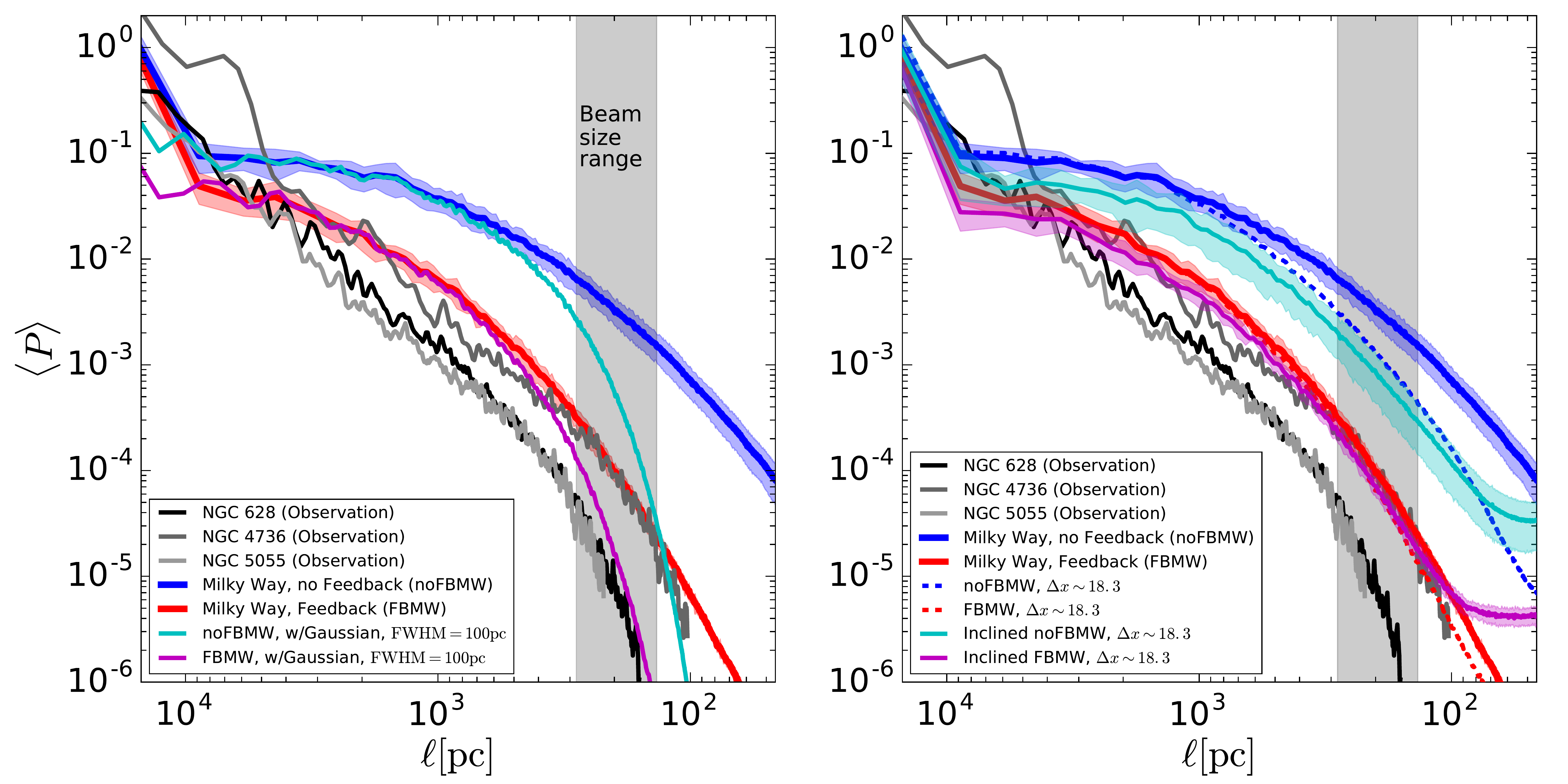}
		\caption{
		Direct comparison of \HI{} surface density power spectra of NGC 628, 4736 and 5055 (black, dark grey and grey solid lines respectively) from our THINGS sample and our Milky Way simulation (red and blue solid lines). The grey shaded region indicates the range of beam sizes given for the this subset of \thingstext. Left panel: We show the effect of first convolving the surface density map of the feedback (magenta) and no feedback (cyan) runs with a Gaussian (FWHM = 100 pc). Right Panel: Time averaged power spectra of the simulations when analysed at $\Delta x \sim18.3\pc$ (dashed lines) and $\Delta x \sim4.6\pc$ (solid lines). Over-plotted are the time averaged spectra of the simulations inclined at $40^\circ$, with feedback (magenta) and without feedback (cyan) at a resolution of $\Delta x \sim18.3\pc$.
		} 
		\label{fig:All_Obs_sim_PS_comp}
\end{figure*}

%--------------------------------------------------------------------------- Section: Results: Surface Density PS -------------------------------------------------------------
\subsection{Structure of gas density field}
\label{sect:DPS}

%--------------------------------------------------------------------------- Section: Results: Surface Density PS: Observations ------------------------------------------
\subsubsection{Observations}
Fig.~\ref{fig:DPS} shows the surface density power spectra, $\Pk$, computed as described in \S\ref{sect:meth_dpow} with $k=2\pi/\ell$, for the 6 galaxies presented in Table~\ref{table:things}, together with fitted power law exponents $\alpha$, defined by $\Pk\propto k^{-\alpha}$, as well as the spatial resolution limit. The surface density maps for each galaxy are normalised by the total \HI{} gas mass ($M_{\rm H{\tiny I}}$)  of the galaxy before we calculate power spectra.

We find that the \HI{} power spectra in all galaxies are all well represented by single power laws over a wide spatial range of a few $100\pc$ up to $\sim 10\kpc$, in agreement with previous studies \citep[e.g.][]{Walker:2014aa}. The power law exponents are found to be in the range $1.6 \lesssim \alpha \lesssim 2.8$. On scales $\lesssim$ few $100\pc$ the spectra steepens, a feature often argued to be an observational signature of the thickness of the gas disc \citep[e.g.][]{Dutta:2009ab}. Indeed, previous work \citep[e.g.][]{Elmegreen:2001aa,Padoan:2001aa,Dutta:2009ab,Zhang:2012aa} have attributed a break in power spectrum to the scale height ($h$), marking a transition from large scale 2D turbulence to 3D on scales $\ell\lesssim h$ \citep[but see][]{Combes:2012aa}.  However, we find that the breaks in the \HI{} spectra always coincide with the resolution limit of \thingstext, making it difficult to make robust claims.

%--------------------------------------------------------------------------- Section: Results: Surface Density PS: Simulations --------------------------------------------
\subsubsection{Simulations}
\label{sect:SDPS}

In Fig.~\ref{fig:PS_Density} we show the resulting \HI{} power spectra from the simulated galaxies, computed as described in \S\,\ref{sect:meth_dpow}. The solid lines are time-average spectra over a period of 100 Myr (separated by $\Delta t=25\Myr$), starting at $t=350$ Myr for all simulations, where the shaded regions show the associated 1$\sigma$ spread in the time averaged data. We have compared our models presented here to lower resolution simulations ($\Delta x\sim18.3\pc$) and found that the power spectra steepen on scales $<10\Delta x$ \cite[see also][]{Joung:2006aa}. We therefore only present results on scales $>10\Delta x$ ($\gtrsim 46$pc).

The simulated \HI{} spectra differ markedly in the simulations with and without stellar feedback where all three different galaxy models follow the same trend; $\Pk$ in models without feedback are more shallow and feature more power on small scales compared to their feedback counterparts, which feature steeper spectra on scales $\lesssim 1$kpc. The increase in small scale power in the models without feedback leads to a decrease in $\Pk$ on large scales ($\gtrsim 1$kpc), as expected from fragmentation. Interestingly, we find a very small scatter between the analysed simulation snapshots over time, indicating that over an orbital time, the density field is roughly in `steady state'.

On scales $\ell\lesssim1\kpc$ the power law fits to the LMC simulation are in good agreement with observations,  \citep{Elmegreen:2001aa,Block:2010aa}. Indeed, \cite{Block:2010aa} found that power spectra of LMC emission at 24, 70, and 160 $\mu m$ have a two-component power-law structure with a shallow slope of $1.6$ for $\ell\gtrsim 100\pc$ , and a steep slope of $2.9$ on smaller scales, in excellent agreement with our LMC simulation including feedback, but in stark contrast to the model without.

\cite{Stanimirovic:1999aa} calculated an $\alpha=2.85$ for the SMC, while \cite{Pilkington:2011aa} found $\alpha=3.2$, on all scales $\ell<7\kpc$. Such steep spectra are not recovered in our models, except for $\ell<100\pc$. Our SMC model with feedback is quite well fit by a single power-law on scales $100<\ell<7\kpc$, in agreement with observations, but with a shallower slope. It is possible that the differences here arise due to the definition of $\Pk$, e.g. being angle-averaged or not, which we leave for a future investigation. Furthermore, the lack of a cosmological environment may affect our results, as the LMC and SMC are passing through the Milky Way halo and experience strong tidal forces. Such forces produce large scale structures such as the Magellanic  Bridge and Magellanic Stream, which would increase the power on large scales.

Finally, we note that all measured \HI{} spectra steepen on small scales, with a break around $\ell_{\rm break}\sim 100-200\pc$ for all simulated galaxies featuring feedback. This scale is resolved in all models, and coincides with the thickness of the \HI{} layer, in agreement with the analysis of \cite{Dutta:2009ab}, as discussed above.

\begin{figure}
	\includegraphics[width=0.45\textwidth]{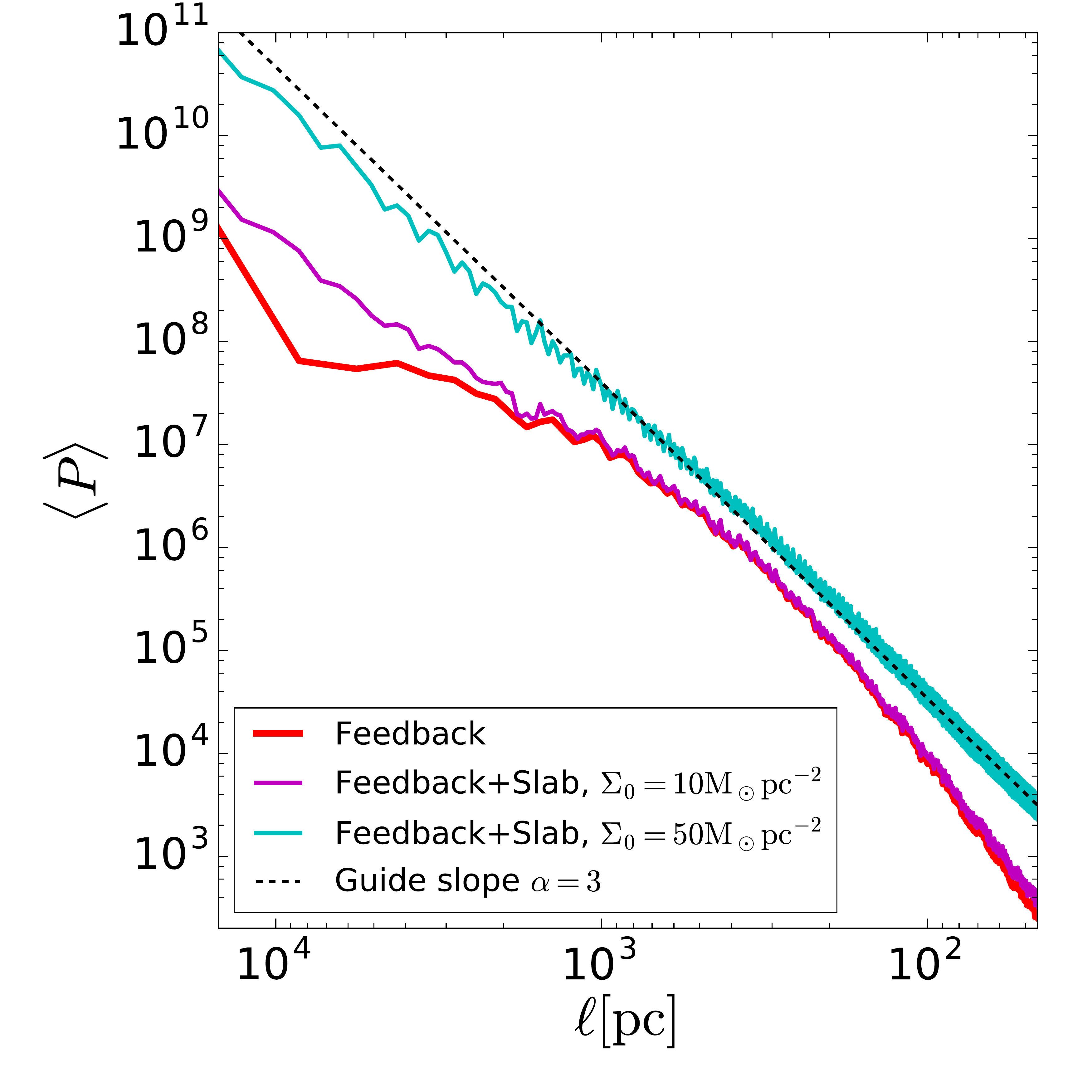}
	\caption{
		Summary of Slab Experiments. We show the power spectra of our Milky Way feedback simulation (FBMW) at $t=400$ Myr with the addition of a uniform slab which is added to the data before calculating the power spectrum. Shown are two slabs with different surface densities $\Sigma_0=10M_\odot \pc^{-2}$ (magenta line), $\Sigma_0=50M_\odot \pc^{-2}$ (cyan line) and the simulation without slab (red line). The dashed (black) line shows a power law slope of $alpha=3$. The spectra are all normalised via their half-light radius (see \S\ref{sect:den_smooth}).
		}
		\label{fig:PS_Dcuts}
\end{figure}

%--------------------------------------------------------------------------- Section: Results: Surface Density PS: Comparison -------------------------------------------
\subsubsection{Direct comparison}
\label{sect:den_smooth}

In the left panel of Fig.~\ref{fig:All_Obs_sim_PS_comp} we compare $\Pk$ from the simulated Milky Way-like galaxy directly to a subset of the \things sample (NGC 628, 4736 and 5055). This subset was selected to compare a range of masses and inclination angles to our simulated galaxies. To compare the relative power and shape of the spectra we normalise the surface density map of each galaxy (simulation and observation) by the \HI{} mass within 1 half-light radius ($r_{1/2}$) from the galactic centre. For NGC 628, 4736 and 5055 this corresponds to $r_{1/2}=5$, $1.75$ and $4.7\kpc$ respectively \citep[values from ][]{Belley:1992aa,Martin:1997aa,Thornley:aa} while for our Milky Way simulations this corresponds $r_{1/2}=3.1$-$3.2\kpc$, depending on the snapshot. 

The feedback model is in good agreement with observations in terms of relative power on all scales, and is an especially good match to NGC4736, which like our simulations is more compact than NGC 628 or 5055, and features a factor of a few less power on scales $\ell\lesssim 5\kpc$. The fact that stellar feedback can affect the distribution of \HI{} up the several kpc, rather than typical scales of SNe bubbles ($<100 \pc$) is dramatic, and illustrates its importance for galaxy evolution.

As discussed above, $\Pk$ from observations are better fit by a single power-law compared to the simulations. This is not necessarily due to the effect of stellar feedback, but rather the presence of a more extended \HI{} disc that is not present in our initial conditions. Indeed, extended low density \HI{} discs and structures are observed out to large galactic radii ($R\sim100\kpc$) \citep{Oosterloo:2007aa,Scott:2014aa}. \cite{Bigiel:2008aa} found extended \HI{} distributions around several of the \things spirals and dwarf irregulars at a nearly constant $\Sigma\sim10 \Msol \pc^{-2}$ out to at least one optical radius ($r_{25}$). A more extreme example is the compact dwarf galaxy NGC 2915 which is surrounded by a very extended \HI{} disc ($r\gg r_{25}$), unable to forms stars \citep{Meurer:1994aa,Meurer:1996aa}. The source of these extended regions is still not clear, but is currently thought to be a remnant of cosmological accretion or the result of mergers and interactions between galaxies \citep[][and reference within]{Oosterloo:2007aa}.  Our simulations use an exponential surface density profiles for the initial conditions, which more closely resembles the observed stellar and \HH distributions. To investigate wether the lack of large scale \HI{} in our simulations can be the cause of the difference between observations and simulation at large scales, we add a low surface density extended distribution of gas to our simulated galaxies before calculating the power spectra. 

In Fig.~\ref{fig:PS_Dcuts} we show the results for a uniform density distribution of gas ($\Sigma(r)=\Sigma_0$) being added to all cells, with $\Sigma_0=10$ or $50\Msol \pc^{-2}$. In both cases the added gas boosts the power on large scales, leading to a steepening of the spectra. In the case of $\Sigma_0=50\Msol \pc^{-2}$, $\Pk$ becomes relatively well fit by a single power-law, in good agreement with observations, albeit with a larger power-law exponent than observed ($\alpha\sim3$). We note that the analytical expectation\footnote{the Fourier transform (${\tilde f}(k)$) of uniform disc is given by ${\tilde f}(k)\propto\frac{J_1(ak)}{k}$, where $J_1$ is the Bessel function of the first kind of order $1$. In the regime of this work $ak>>1$ and therefore $J_1\sim(ak)^{-1/2}$. ${\tilde f}(k)$ then becomes proportional to $k^{(-3/2)}$ making the power spectrum proportional to $k^{-3}$ } for the power spectrum of a uniform disc is indeed $P(k)\propto k^{-3}$. In the case of more complex structure, such as galaxies, the power spectrum is a superposition of the uniform disc plus the signal from the `fine' structure of the galaxy i.e. GMCs and spiral arms. $\Sigma\ge 50\Msol \pc^{-2}$ is an unusually high \HI{} surface density at large galactic radii \citep[][]{Bigiel:2008aa}, and should be considered an extreme case, but illustrates that missing extended gas will affect the shape of $\Pk$.

Finally, both observational and simulated data contain smearing of the signal on the smallest scales due to limited resolution, which could affect analysis of such data. We briefly explore this effect on the observed density power spectra by convolving the surface density maps of our Milky Way simulations with a Gaussian prior to calculating the power spectra. Fig.~\ref{fig:All_Obs_sim_PS_comp} shows the results for a Gaussian with a Full Width Half Maximum (FWHM) equal to $100\pc$ on $\Pk$\footnote{We follow the standard definition of a Gaussian, $\phi(x)=ae^{-(x-b)^2/(2c^2)}$, where is $a$ normalising constant, $b$ is the position of the peak of the Gaussian and $c={\rm FWHM} / (2\sqrt{2\ln{2}})$.}. We find that this procedure removes power on scales less than $\sim3\,$-$\,4\times$ of the FWHM, hence affecting the overall shapes of the spectra. 
This effect can be dramatic, but does not lead to difficulties in disentangling the effect of stellar feedback (when the spectra is appropriately weighted). In our models the overall match to observations in the absence of feedback is always poor, regardless of choice of smoothing scale, indicating that this effect is subdominant for the {\small THINGS} sample.

%--------------------------------------------------------------------------- Section: Results: Surface Density PS: Inclination Effects ------------------------------------
\subsubsection{Effect of inclination}
\label{sect:den_inc}

In the previous sections we considered all observed galaxies as face on, i.e. $i=0\degree$. However as shown in Table~\ref{table:things} this is not the case for our \things sample. We explore the effect of inclination of the galaxy on the power spectrum by observing our simulated galaxies at a $i=40\degree$ inclination, which is the mean inclination in our \things sample. For computational feasibility we analyse the simulation at a resolution of $\Delta x\sim18.3\pc$ for both the density power spectra and the 2D kinetic energy spectra. We note that reducing the resolution of the analysis has the effect of removing small scale power from the no feedback simulations, but very little effect on the feedback runs.

The right panel of Fig.~\ref{fig:All_Obs_sim_PS_comp} shows the effect of inclining the galaxy on the surface density power spectrum; compared to the face-on case only a small increase in power on scales of small scales ($\ell\lesssim100\pc$) in both feedback and no feedback runs. This is expected due the increase in column density introduced by inclining the galaxy, but on larger scales we find very little change in the shape of the spectra in both cases. From these results we conclude that the effect of inclination on a density power spectrum is to increase power on small scales, and leave large scale unaffected, making it possible to `observe' the effect of feedback directly without inclination correcting.

\begin{figure*}
	\begin{center}
		\includegraphics[width=0.75\textwidth]{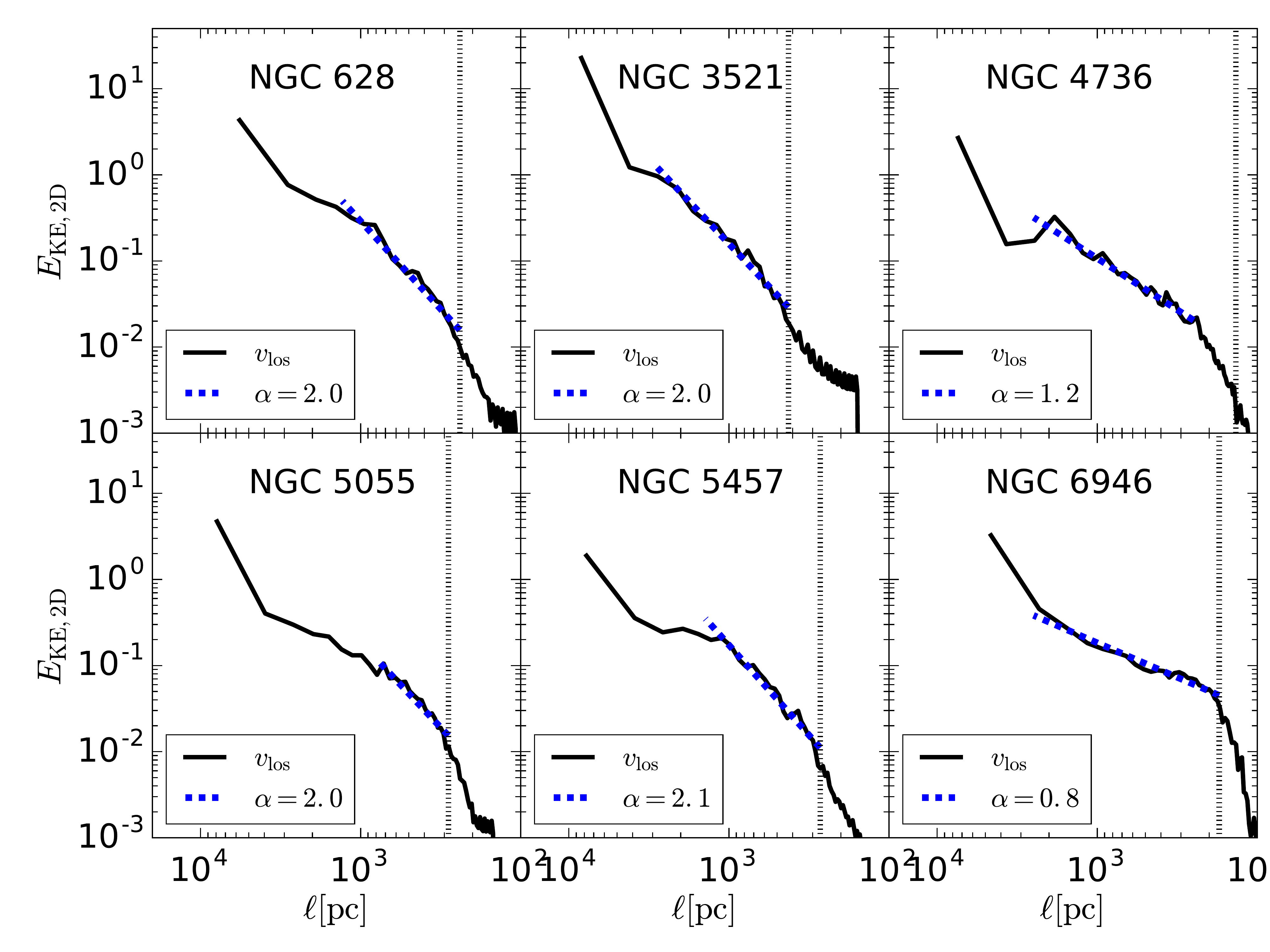}
		\caption{2D kinetic energy ($\Sigma_{\rm HI}^{1/2}v_{\rm los}$) power spectra for our \things sample (black solid lines). The blue dashed line shows a fitted power law to the spectra, with the gradient of the slope given in legends. The vertical black dotted line shows the beam width for each galaxy (see Table \ref{table:things} for values). The $E(k)$ presented here have all been normalised (see \S\ref{sect:vel_smooth} for details).	
		}
		\label{fig:KEPS_things}
	\end{center}
\end{figure*}
\label{sect:KEPS_sims}

\begin{figure*}
	\begin{center}
		\includegraphics[width=0.9\textwidth]{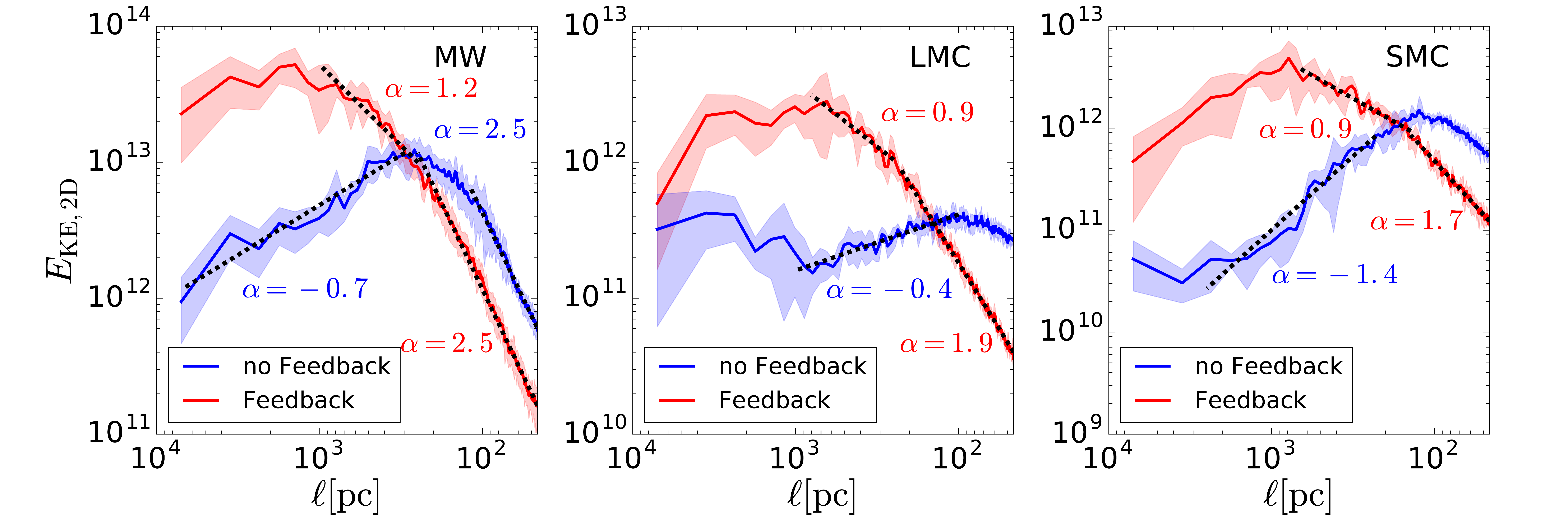}
		\caption{
		2D kinetic energy ($\Sigma_{\rm HI}^{1/2}v_{\rm los}$) time-averaged power spectra for our simulations. From left to right we show the resulting spectra from the MW, LMC and SMC models with feedback shown in red and no feedback shown in blue. The dashed black lines show our fitted power law with gradients next to each fit. The shaded regions around each power spectrum show the $1\sigma$ deviation from the mean. The $E(k)$ presented here have \emph{not} been normalised.
		}
		\label{fig:PS_2D_KE}
	\end{center}
\end{figure*}

%--------------------------------------------------------------------------- Section: Results: Energy Spectra ------------------------------------------------------------------
\subsection{Kinetic energy of the interstellar medium}
\label{sect:KEPS}
In this section we analyse the impact of feedback on the velocity structure of the simulated and observed ISM. As described in \S\ref{sect:meth_vpow}, we do so by computing the 2D line-of-sight kinetic energy power spectrum ($E_{\rm KE,2D}(k)$) for the simulated and observed galaxies, as well as full 3D spectra ($E_{\rm KE,3D}(k)$) for the simulations, allowing us to compare the results to well known expectations of turbulent scalings for incompressible Kolmogorov turbulence ($E(k)\propto k^{-5/3}$) and compressible super-sonic turbulence ($E(k)\propto k^{-2}$).

%--------------------------------------------------------------------------- Section: Results: Energy Spectra: Observations -----------------------------------------------
\subsubsection{Observations}
\label{sect:KEPS_things}
In Fig.~\ref{fig:KEPS_things} we show the line-of-sight $E(k)$ from the \things galaxies. Four of the galaxies in our sample (NGC  628, 5055, 5457 and 3521) feature steep power spectra on scales above the resolution limit, with $E(k)\propto k^{-\alpha}$ and $\alpha\sim 2$, suggesting that the ISM of these galaxies is compressible and supersonic. All but the latter galaxy feature a reduction in steepness in the spectra on scales $0.5\kpc\lesssim \ell\lesssim 1\kpc$, possibly related to disc thickness.

The ISM in the starburst galaxy NGC 4736 is rather different compared to the other galaxies, with a large ring of cold gas and young stars in its central region. This is possibly the origin of the more shallow $E(k)$ measure in this galaxy on $\sim \kpc$ scales, with $\alpha\sim1.2$, indicating that the contribution from small scales is significant. NGC 6946 features an even shallower $E(k)$, with $\alpha\sim0.8$. This galaxy also has a complicated dynamical structure, with three nested bars, a double molecular disc and a nuclear starburst region \citep{Romeo:2015aa}. The velocity map of NGC 6946 \citep[see Fig.~65, bottom left panel, of][]{THINGSpaper} shows large structures, with sizes of $\sim5\kpc$ or greater, with little variation ($\Delta v\lesssim25 {\rm kms^{-1}}$) in velocity. These relatively uniform regions could possibly account for the shallower shape of $E_{\rm KE,2D}(k)$.

\begin{figure*}
	\begin{center}
		\includegraphics[width=0.9\textwidth]{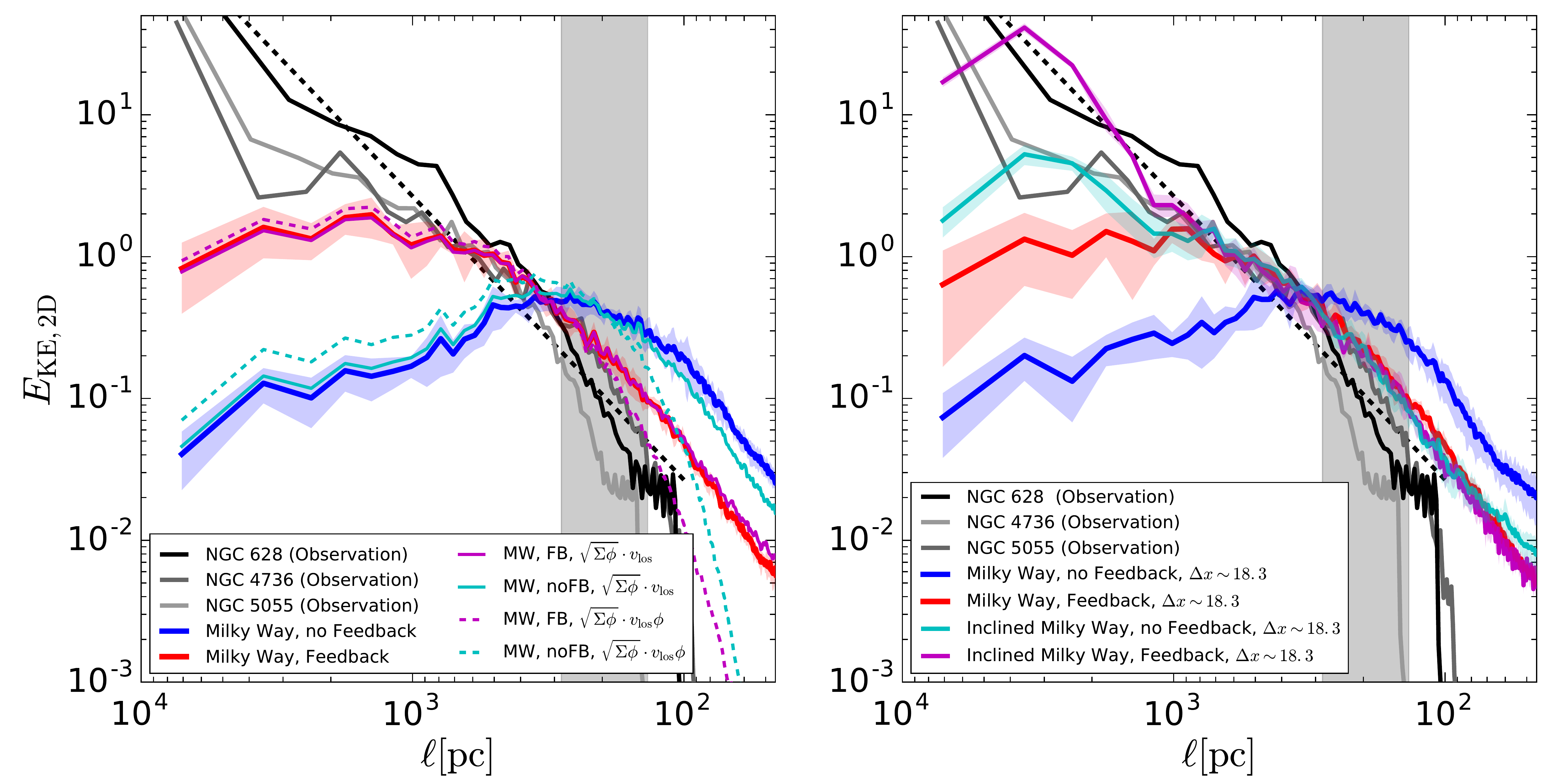}
		\caption{Direct comparison of the 2D kinetic energy ($\Sigma_{\rm HI}^{1/2}v_{\rm los}$) power spectra of NGC 628, 4736 and 5055 (black, dark grey and grey solid Lines respectively) from our \things sample and our Milky Way simulation (red and blue solid lines). The grey shaded region indicates the range of beam sizes given for the this subset of \things. Left: Shown in magenta (feedback) and cyan (no feedback) are the results of first convolving the surface density map (solid lines) and the surface density map as well as the velocity map (dashed lines) with a Gaussian (FWHM=100pc). Right: Inclination test. Time averaged spectra of the simulations inclined at $40^\circ$ are shown in magenta (feedback) and cyan (no feedback). All simulation spectra in this panel are calculated at a resolution of $\Delta x\sim18.3\pc$. These spectra are all normalised, see \S\ref{sect:KE_sims} for details. 
		}
		\label{fig:KE_things_sim_comp}
	\end{center}
\end{figure*}
%reworked

%--------------------------------------------------------------------------- Section: Results: Energy Spectra: Simulations -------------------------------------------------
\subsubsection{Simulations}
\label{sect:KE_sims}

In Fig.~\ref{fig:PS_2D_KE} we show $E_{\rm KE,2D}(k)$ for all simulations, using only the line-of-sight velocity field as observed face-on. As for the density spectra, we show the time-averaged spectra of the feedback/no feedback simulations for the MW, LMC and SMC simulations respectively, and we only show result down to 10 resolution elements ($10\Delta x$). 

We again find a dramatic difference between the feedback and no feedback simulations; feedback regulation results in steep power spectra, with $\alpha\sim 2-3$ on scales $\lesssim 0.5 \kpc$, with a transition into shallower relations on large scales, in good agreement with the a subset of the observations discussed above. Most power in the feedback run is hence present on large ($\kpc$) scales, with a cascade to small scales. Without feedback, $E_{\rm KE,2D}(k)$ \emph{increases} down to scales of a few $100\pc$ for all simulated galaxies, indicating the presence of dense cloud structures, and turns over on smaller scales. Large scale turbulence imparted by large scale rotation and gravitational instabilities is no longer present, as power has cascade once into dense star forming clouds where it is `locked up' instead of being disrupted and returned to the large scale driving. We will explore these concepts more in future work (Agertz et al. in prep).

%--------------------------------------------------------------------------- Section: Results: Energy Spectra: Comparison -------------------------------------------------
\subsubsection{Direct comparison}
\label{sect:vel_smooth}

In Fig.~\ref{fig:KE_things_sim_comp} we directly compare\footnote{To enable this direct comparison we normalise $E_{\rm KE,2D}(k)$ by enforcing $\int_{\ell=0.1\kpc}^{\ell=1\kpc} E(k)\mathrm{d}k=1$. 
} $E_{\rm KE,2D}(k)$ from the Milky Way-like simulations to a subset of \things sample (NGC 628, 4736 and 5055). Without feedback we find an excess of power on small scales ($\ell\lesssim 200-300\pc$), and a significant lack of power on large scale with a peak in the spectrum on intermediate scales. However, the feedback runs provide a better match to observations on all scales. This highlights the role of stellar feedback, together with gravity and shear, in regulating the energetics of the ISM on all scales. $E_{\rm KE,2D}(k)$ from the feedback regulated simulations are a better match to observations for $\ell<1\kpc$, indicating that our adopted feedback model can readily predict the scale dependence of the (line-of-sight) kinetic energy field of real galaxies on small scales. 

An even closer agreement on sub-kpc scales is found when smoothing the simulated data, as show in Fig.~\ref{fig:KE_things_sim_comp}. Smoothing\footnote{We explore two regimes of smoothing: applying a Gaussian ($\phi$) only to the surface density map ($\sqrt{\Sigma_{\rm HI}\phi}\cdot v_{\rm los}$) and applying the Gaussian to both the surface density map and line of sight velocity map ($\sqrt{\rho\phi}\cdot v_{\rm los}\phi$)}, again using a Gaussian with a ${\rm FWHM} = 100\pc$, predominantly gives rise to less small scale power which brings the feedback model to overlap with observations. The model without feedback is affected in a similar way, but is still inconsistent with the observations on all scales.

What about the observed large scale power? In the right panel of Fig.~\ref{fig:KE_things_sim_comp} shows the effect of inclination of $E_{\rm KE,2D}(k)$, carried out in the same manner as in \S\ref{sect:den_inc}. When inclined we find that the non-turbulent rotational velocity dominates the entire signal, which brings both the feedback and no feedback model into close agreement with observations. For line-of-sight energy spectra to be able to be able to differentiate feedback models, careful subtraction of the gas rotational velocity is needed, e.g. using ``tilted-ring model'' \citep{Rogstad:1974aa} or 3D analogues such as the method of \cite{Teodoro:2015aa}, as used to model \HI{} rotation curves of dwarf galaxies \citep{Read:2016aa, Read:2016ac}. Such a analysis is beyond the scope of the this work.

\begin{figure*}
	\begin{center}
		\includegraphics[width=1.\textwidth]{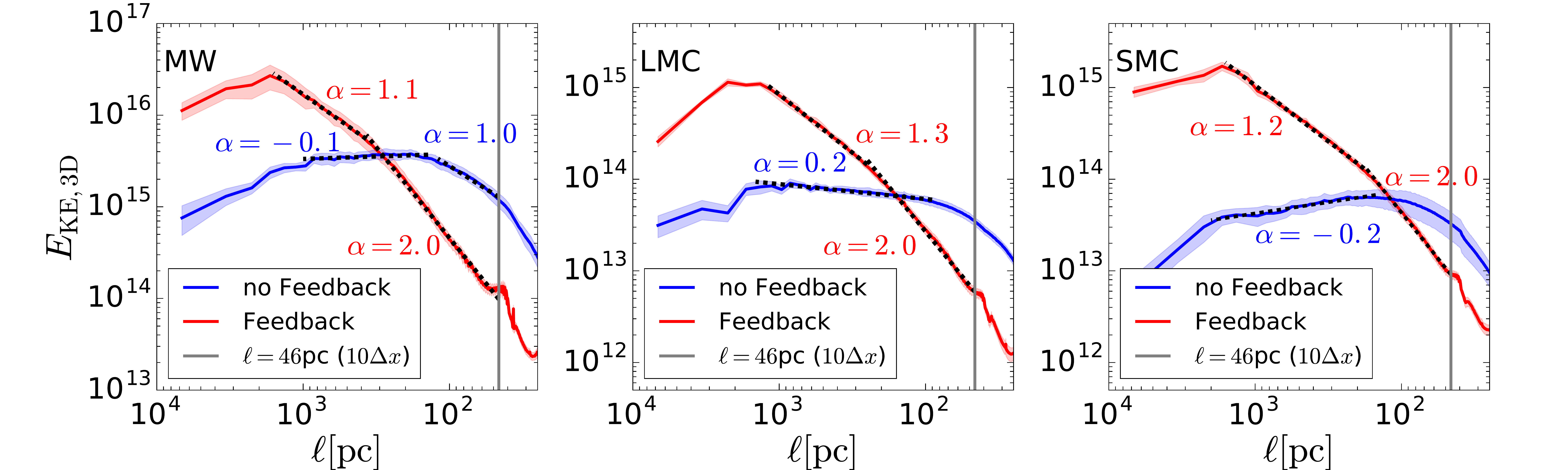}
		\caption{
		Full 3D kinetic energy ($\rho_{\rm all}^{1/2}\bvect{v}$) power spectra spectra for each galaxy (Milky Way left, LMC centre and SMC right) using the full 3D velocity field ($\bvect{v}$). Each panel shows the time-averaged spectra (solid colour lines) and the $1\sigma$ deviation from the time-average (shaded colour region).The black (dashed) lines represent various gradients which are detailed on the panel. Red represents feedback simulations and blue no feedback simulations in all panels. The minimum trusted scales is shown with the vertical black line.
		}
		\label{fig:PS_KE}
	\end{center}
\end{figure*}

\begin{figure}
	\begin{center}
		\includegraphics[width=0.4\textwidth]{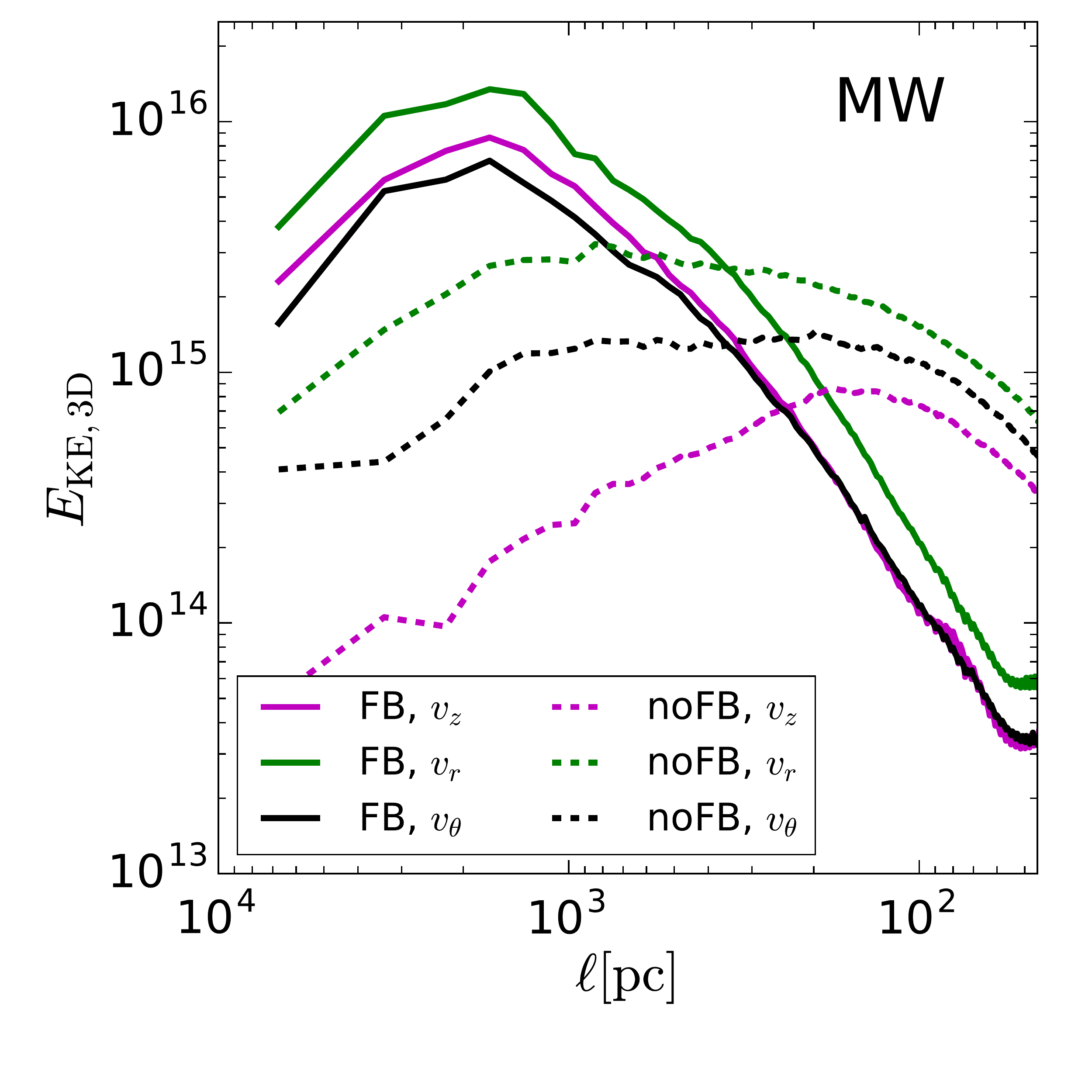}
		\caption{
		Time averaged 3D kinetic energy ($\rho_{\rm all}^{1/2}v_{\rm c}$) power spectra spectra for each velocity component of the Milky Way simulations. We show the vertical ($v_z $, magenta), radial ($v_r$,  green) and tangental ($v_\theta$, black) velocity components. We distinguish between feedback and no feedback with solid and dashed lines respectively. We note that we have not normalised, i.e. they show the relative power of each component
		}
		\label{fig:PS_KE_comp}
	\end{center}
\end{figure}

\begin{figure*}
	\begin{center}
		\includegraphics[width=0.7\textwidth]{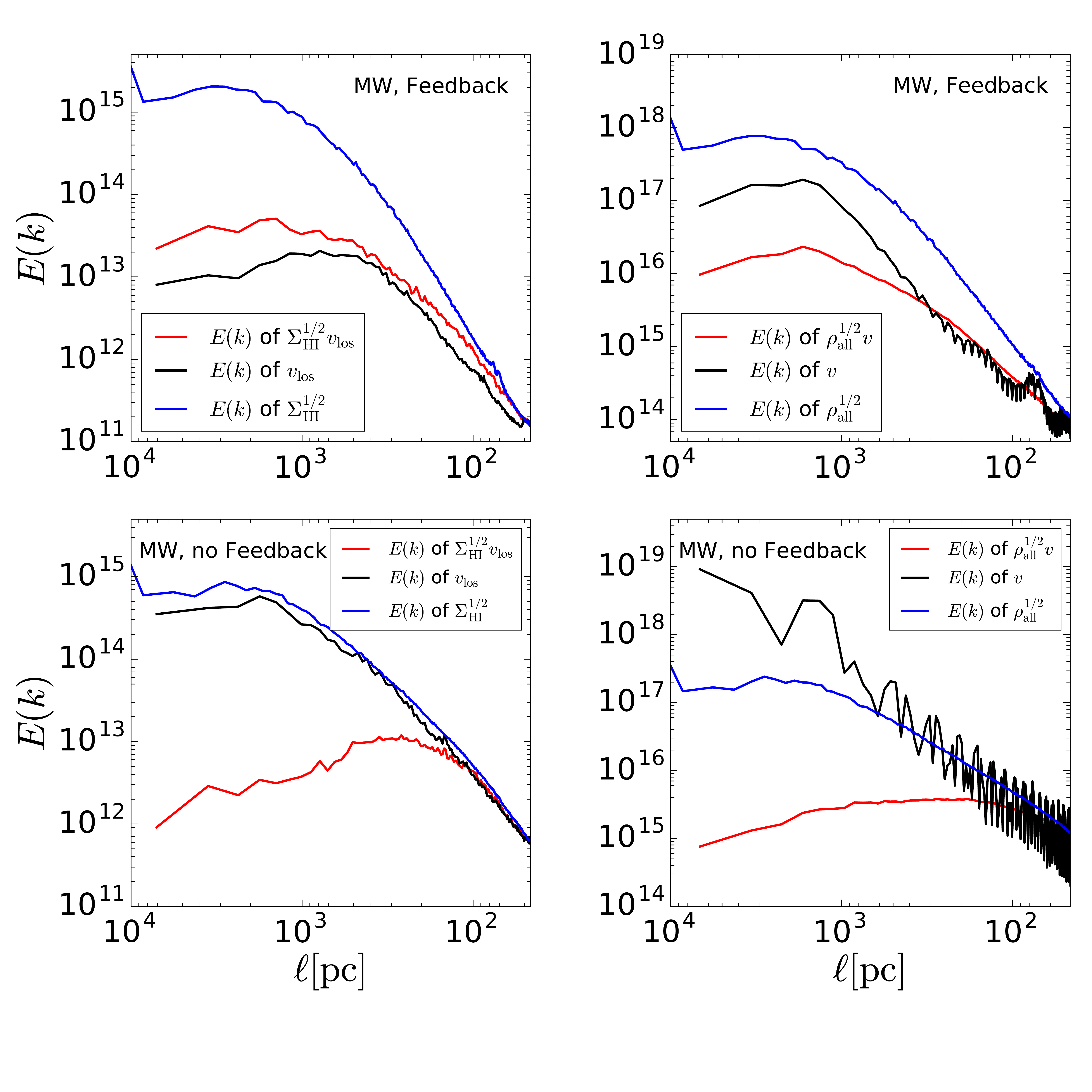}
		\caption{
		Comparison of the spectra for $\Sigma_{\rm HI}^{1/2}v_{\rm los}$, $v_{\rm los}$ and $\Sigma_{\rm HI}^{1/2}$ (left panels) and $\rho_{\rm all}^{1/2}\bvect{v}$, $\bvect{v}$ and $\rho_{\rm all}^{1/2}$ (right panels). Top: Comparison of simulation including feedback. Bottom: Comparison of simulations neglecting feedback. The spectra are normalised to the value of $E(k)$ for $\Sigma_{\rm HI}^{1/2}v_{\rm los}$ (left panels) and for $\rho_{\rm all}^{1/2}\bvect{v}$ (right panels) at $\ell=46\pc$.
		}
		\label{fig:noise_check}
	\end{center}
\end{figure*} 

\begin{figure*}
	\begin{center}
		\includegraphics[width=0.8\textwidth]{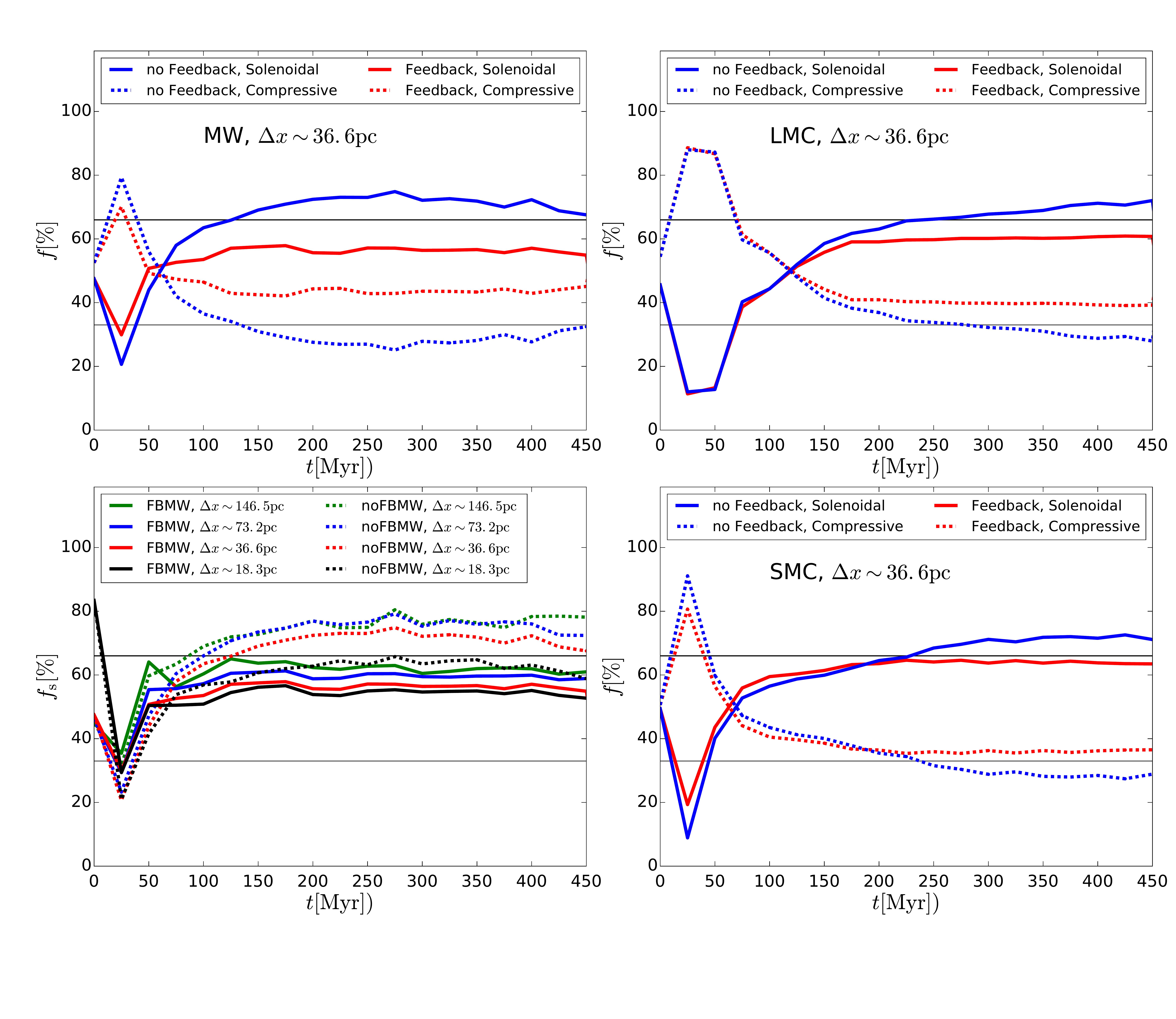}
		\caption{
		Top row and bottom right: Mass weighted fraction of turbulent kinetic energy in solenoidal  (solid lines) and compressive (dash lines) turbulent motion. Each panel show the data for one set of simulations: the panel on the top left for the Milky Way-like, the top right panel shows the LMC-like and the bottom right panel the SMC-like simulations. Simulations with feedback are shown with red lines while those without are shown with blue. We indicate the equipartition fractions with horizontal black ($f=66\%$) and grey ($f=33\%$) lines.
		Bottom left: Comparison of the solenoidal fraction of our Milky Way-like simulation when calculated at different resolutions (see legend for resolution). Solid lines show results for the feedback run and dashed lines for no feedback. }
		\label{fig:Energer_frac}
	\end{center}
\end{figure*}

%--------------------------------------------------------------------------- Section: Results: 3D Energy Spectra -------------------------------------------------------------
\subsection{3D energy spectra}
\label{sect:3D_energy_spec}

We now turn to the full 3D kinetic energy $E(k)$ (see \S\ref{sect:meth_vpow}), which we show in Fig.~\ref{fig:PS_KE} for all three galaxy sizes. As for the 2D case, the feedback simulations feature a much steeper power spectra compares to their no feedback counterparts. Without feedback power cascades to, and remains, on small scales, hence creating shallow energy spectrum with a steepening occurring on scales of individual clouds. In all three galaxy models, feedback maintains power law slopes of almost exactly $\alpha=2$ on scales $\lesssim$ several 100 pc, as expected in super-sonically turbulent flows. On scales $\gtrsim$ few 100 pc a more shallow scaling of $E_{\rm KE,3D}(k)$ is measured, possibly indicating the thickness of the simulated discs. The spectra feature very little scatter over time, indicating that turbulence driven by stellar feedback and, coupled with the large scale driving, has reached steady-state over the time scale that we carry out analysis. 

These results complement previous results in the literature. For example, \cite{Joung:2006aa}, \cite{Martizzi:2016aa} and \cite{Padoan:2016aa} modeled SNe driven turbulence in isolated small scale boxes ($<1 \kpc$) and measured $E(k)$, finding that power one small scales ($\ell\lesssim100\pc$) follows $E(k)\propto k^{-2}$. On larger scales, their energy spectra becomes shallower and feature significantly less power than what we predict here, illustrating that full galactic models are necessary to account for the transfer of energy between scales inside of galaxies. In future work we will explore the driving scale of the $E(k)$ and the interaction between small scale driving (stellar feedback) and large scale driving (gravity).

We have also performed the same analysis on the three velocity components in a cylindrical coordinate system ($v_z$, $v_r$ and $v_\theta$) for the simulated Milky Way galaxy, and find close to identical scalings for each component as for the total kinetic energy field in the feedback simulation, see Fig.~\ref{fig:PS_KE_comp}, with the radial component containing most of the power. In the no feedback case we see a significant difference between the three components, with $E_{v_z}<E_{v_\theta}<E_{v_r}$ on all scales. The lack of feedback here leads to little power in the vertical component, being $\sim 2$ dex lower compared to its feedback counterpart. 

\cite{Bournaud:2010aa} found, using high resolution models of an LMC-like galaxy, that $E(k)$ of $v_r$ and $v_\theta$ are almost identical, while $v_z$ follows shallower power law up to scales of $\ell\sim200-300\pc$ before levelling off. The reason for this discrepancy could be due to the nature of their feedback model, e.g. the lack of pre-SNe feedback limits the effectiveness of the SNe for not just the density power spectra (see \S\ref{sect:dis:sim2sim}) but also for $E(k)$.

%--------------------------------------------------------------------------- Section: Results: Noise vs signal -------------------------------------------------------------------
\subsection{Velocity or density fluctuations?}\label{sect:velnoise}
It is unclear from the above analysis what gives rise to the measured energy spectra. It is possible that $E(k)$ for our simulations is driven by fluctuations in the density field only i.e. $E(k)$, as defined here, is just a measure of $\sqrt{\Sigma_{\rm HI}}$ (2D) or $\sqrt{\rho_{\rm all}}$ (3D), with random fluctuations from the velocity component superimposed onto the spectra. To test this possibility we compare the spectra of $\Sigma_{\rm HI}^{1/2}v_{\rm los}$, $v_{\rm los}$ and $\Sigma_{\rm HI}^{1/2}$ in 2D and $\rho_{\rm all}^{1/2}\bvect{v}$, $\bvect{v}$ and $\rho_{\rm all}^{1/2}$ in 3D for our Milky Way-like simulation in Fig.~\ref{fig:noise_check}. Henceforth we refer to these quantities as $E_{\rm KE,2D}(k)$, $E_{\rm v_{\rm los}}(k)$, $E_{\rm \Sigma}(k)$, $E_{\rm KE,3D}(k)$, $E_{\bvect{v}}(k)$ and $E_{\rm \rho}(k)$ respectively. Note that in order to facilitate a comparison of the shape of the spectra, we normalise them to the same power at $\ell=46$ pc. 

For the feedback model we find that $E_{\rm KE,2D}(k)$ and $E_{\rm v_{\rm los}}(k)$ are very similar, indicating that the energy spectrum is shaped by the velocity fluctuations. In the 3D feedback case we find that on scales less than a few 100 pc, $E_{\rm KE,3D}(k)$ and $E_{\bvect{v}}(k)$ trace each other, although $E_{\bvect{v}}(k)$ features a lot of noise, while on large scales $E_{\bvect{v}}(k)$ increases in power. This \emph{relative} increase in power may be due to feedback driven outflows of tenuous gas venting out of the disc (see the edge-on view in the bottom left of Fig.~\ref{fig:Maps}). Both $E_{\rm \Sigma}(k)$ and $E_{\rm \rho}(k)$ poorly match the respective velocity and kinetic energy spectra, suggesting the velocity fluctuations play an important role in explaining the kinetic energy spectrum when feedback is present.

Without stellar feedback, $E_{\rm v_{\rm los}}(k)$ and $E_{\rm \Sigma}(k)$ are almost identical, i.e. the velocity structure of the gas appears to trace that of the density. However, combining these two produces a non-linear result, here a turnover at $\ell\sim200-300\pc$, with little power present on large scales. $E_{\bvect{v}}(k)$ and $E_{\rm \rho}(k)$ show a similar behaviour for $\ell\lesssim1\kpc$, with significant noise\footnote{This behaviour likely originates from sharp features in the velocity field, leading to oscillations in the Fourier transformed quantities  \citep[This is the familiar `Gibbs' phenomenon;][]{matmet} } in the velocity component, before diverging at larger $\ell$.

From these comparisons we conclude that the kinetic energy spectra we present for $\Sigma_{\rm HI}^{1/2}v_{\rm los}$ and $\rho_{\rm all}^{1/2}\bvect{v}$ are not driven by the density field alone.

%--------------------------------------------------------------------------- Section: Results: Compression & Solenoidal motions ----------------------------------------
\subsection{Compressive vs. solenoidal motions}
\label{sect:compvssolenoidal}

Having established that energy spectra of the turbulent gas in galaxies with and without stellar feedback differ markedly on all scales, it is interesting to explore how stellar feedback affects the nature of turbulence by computing the fraction of kinetic energy present in compressive\footnote{Compressive modes includes both compression and rarefaction of the gas} (curl-free, $\nabla\times\bvect{v}=\bvect{0}$) and solenoidal\footnote{Some authors refer to solenoidal modes as vorticity.} (divergence-free, $\nabla\cdot\bvect{v}=0$) motions. This is an important characteristic of the ISM, e.g. the shape of the density PDF depends on the type of turbulent forcing, as demonstrated by numerical simulations of super-sonic flows \citep[e.g.][]{Padoan:1997aa,Federrath:2010aa,FederrathKlessen2013}. 

In compression-dominated turbulence, regions with an excess of dense gas widen the PDF from the expected case of isothermal supersonic turbulence of non self-gravitating gas \citep{Vazquez:1994aa, Nordlund:1999aa,Wada:2001aa}. This in turn shifts the median of the PDF to larger densities. For an ISM that has reached equipartition it is expected that $2/3$ of the kinetic energy is in solenoidal motions and 1/3 in compressive \citep[][and references within]{Padoan:1997aa,Kritsuk:2007aa,Federrath:2010aa,Renaud:2013aa,Renaud:2014aa}. The ratio of 2:1 is determined by the dimensionally of each mode, i.e. compression only has one degree of freedom, while solenoidal modes have two.   

\cite{Renaud:2014aa} noted that compression-dominated turbulence is induced by tides during galaxy interactions, and triggers the observed starburst activity over kpc-scale volumes. However, the role of compressive turbulence appears to be less important in more quiescent environments like isolated disc galaxies, where turbulence remains close to equipartition (i.e. solenoidal-dominated regime) \citep[][]{Renaud:2015aa}. There, the Mach number is the main driver of gas over-densities, and thus star formation.

To better understand the nature of the ISM turbulence in our simulations we calculate the fraction of energy in each component via
\begin{equation}
	f_{m} = \sum\limits_{i=1}^n\rho_i\left(\frac{\rho_i v_{m,i}^2}{\rho_i  v_{s,i}^2+\rho_i  v_{c,i}^2}\right)/{\sum\limits_{i=1}^n\rho_i}
	\label{eq:app_fracen2}
\end{equation}
where $m$ refers to the mode (solenoidal or compressive),  $\rho_i$, $v_{s,i}$ and $v_{c,i}$ is the density, solenoidal velocity component  and compressive velocity component of $i^{th}$ cell and $n$ is the total number of cells. We calculate $v_c$ and $v_s$ at a uniform resolution of $\Delta x=36.6\pc$ to ensure that the solenoidal motions are properly captured \citep[i.e. $8\times$ the width of a fully refined cell, as discussed in ][]{Renaud:2015aa}. First derivatives are computed using a stencil of $\pm 4$ cells.
 
In Fig.~\ref{fig:Energer_frac} we show how $f_{\rm s}$ and $f_{\rm c}$ evolves as a function of time in our simulations. After an initial transient phase, the solenoidal modes become the dominate type of kinetic energy in all models, when feedback is not present the solenoidal mode accounts for $\sim 70\%$ of the turbulent energy budget, in good agreement with equipartion expectations. In the models with feedback we find that solenoidal accounts for between 55\% and 66\% of the turbulent energy budget, meaning stellar feedback increases the importance of compressive (shock dominated) turbulences. This is more significant in the larger MW model, whereas the effect is smaller in the dwarf galaxies, being almost negligible in the SMC model.

Despite the strong impact of feedback discussed in the previous sections, the simulations indicate that solenoidal motions dominate the turbulent energy budget. This suggests that energy and momentum injection related to feedback does not lead to compression dominated ISM. \cite{Padoan:2016aa} found that solenoidal motions could be produced by feedback as a result of expanding bubbles interacting with a non-uniform ISM through the baroclinic effect. 

We also note that the fraction of energy in each component depends on scale: as $\Delta x$ increases, and we hence probe gas motions on larger scales, we see an increase (decrease) in the fraction of kinetic energy found in the solenoidal (compressive) mode (see Fig.~\ref{fig:Energer_frac}, bottom left panel). Despite this dependance on scale, $f_s$ is always $>50\%$ and $>66\%$ in feedback and no feedback simulations respectively i.e. independent of scale feedback always increases fraction of energy found in compressive modes.

%-------------------------------------------------------------------------------------------------------------------------------------------------------------------------------------------
%--------------------------------------------------------------------------- Section: Discussion ------------------------------------------------------------------------------------
\section{Discussion}
\label{sec:discussion}

\subsection{Previous results from galaxy simulations}
\label{sect:dis:sim2sim}
A number of studies have analysed the structure of the ISM using power spectra. Both \cite{Combes:2012aa} and \cite{Walker:2014aa} simulated Milky Way-like galaxies including sub-grid models for stellar feedback. In agreement with our findings, both studies found that increasing the strength of feedback resulted in steeper \HI{} density power spectra, fit by single ($P\propto k^{-\alpha}$) or multiple power laws. \cite{Walker:2014aa} found that models with weak feedback (their MUGS suite) featured shallow power-laws ($\alpha\sim 1.2$) similar to $\alpha\sim 1.5$ found in our simulations without feedback. In their MaGICC suite of simulations, featuring significantly stronger feedback (supernovae + `early pre-SN' feedback), their recovered indices of $\alpha\sim 2.5$ closely match ours. However, in contrast to our models, they find that on scales $\ell\lesssim 2\kpc$ their spectra steepen considerably ($\alpha\sim 5$), in contrast to observed galaxies on those scales, as shown in \S\ref{sect:DPS}. The origin of this discrepancy may be that the MaGICC galaxies feature significantly thicker gaseous discs than analysed here, possibly due to the lower spatial resolution in those models ($\sim 155\pc$ compared to the 4.6 pc used here).

\cite{Bournaud:2010aa} modeled an LMC-like galaxy to study the effect of feedback on the structure of the ISM. However they found that their galaxies with and without feedback were both well fit by two power laws, with a break at $\ell\sim 150\pc$, and with almost identical power law indices on large ($\alpha\sim 1.9$) and small ($\alpha_{\rm small}=3.12$) scales. Our LMC model is in excellent agreement with these values (see Fig.~\ref{fig:PS_Density}), but \emph{only} when stellar feedback is present. As for the more massive galaxy models, we found that neglecting feedback resulted in shallower \HI{} spectra, in contrast to the finding on \cite{Bournaud:2010aa}. It is unclear why our models give rise to such different conclusions, but we note that they adopt an equation of state for their ISM, instead of solving the full energy equation, which may stabilise the model without feedback enough to give rise to a structure compatible with their model including stellar feedback. Indeed, a visual inspection of their simulations \citep[see figure 3 and 4 in][]{Bournaud:2010aa} reveals only small differences, mostly a slightly more porous ISM on small scales, with a large scale morphology that does not resemble our fragmented LMC model when feedback is not present.

\cite{Pilkington:2011aa} analysed dwarf galaxies formed in a cosmological context, and argued that single power law fits to their models agreed with observations of similar mass galaxies such as the SMC. However, whereas the SMC indeed is well fit by a single power law, their simulations are better fit by multiple power laws with a break at $\sim450\pc$. This is closer to what we find for our LMC and SMC simulations that include feedback, although the break occurs on smaller scales, and the exact value of the power-law indices in all of our models are larger, likely due to the definition of $\Pk$, see discussion below. 

\cite{Krumholz:2016aa} found that analytical models of feedback-driven turbulence predict a lower velocity dispersion for galaxies with a ${\rm SFR}>1\Msol\,{\rm yr}^{-1}$ compared to gravity-driven only models (see their figure 1). By comparing these models to observational data of gas velocity dispersion as a function of SFR, they argued that gravity is the primary source of turbulence in the ISM on scales typical of gravitational instabilities in galactic discs, i.e. supporting the conclusions of \cite{Goldbaum:2016aa}. Our simulations confirm this notion, where gravity only models feature more kinetic energy than feedback driven models on small scales ($\ell < h$). However, we emphasise that the role of feedback as a complementary driver of turbulence varies strongly with scale (as seen in Fig.~\ref{fig:PS_KE}), with large scale power being of greater importance compared to gravity only models.

We note however several caveats to their model that affect the interpretation of their results. The data set used to distinguish between gravity and feedback models have not been corrected for observational effects, difference in observational method or removal of rotational velocity. Instead, the raw observational data is used as reported by the authors.  One example is the data from \cite{Lehnert:2013aa} which used H$\alpha$ lines to determine the velocity dispersion. These values are only corrected for resolution effects i.e. rotation is not accounted for. Therefore any results derived from these velocity dispersions are most likely an overestimate. 

Furthermore, the models presented in \cite{Krumholz:2016aa} rely on the Toomre stability parameter for gas ($Q_g$), stars ($Q_*$) and  the galaxy disc ($Q$), where $Q^{-1} = Q_*^{-1} + Q_g^{-1}$ \citep[see][]{Toomre:1964aa,Wang:1994aa}, assuming $Q$ and $Q_g$ to be equal to unity for the gravity and feedback models respectively.  We note that $Q=1$ is a strong assumption. In fact, in the \things sample, star-forming spirals feature $Q\sim$ 1--5 \citep[see fig.~4 of ][]{Romeo:2013aa}, $Q_g$ spans an even wider range of values \citep[see fig.~5 of ][]{Romeo:2011aa}, and both $Q$ and $Q_g$ depend on the scale over which they are measured \citep{Hoffmann:2012aa,Agertz:2015aa}. All of the above can significantly affect the conclusions of \cite{Krumholz:2016aa}. 

%--------------------------------------------------------------------------- Section: Discussion: Previous Observations -----------------------------------------------------
\subsection{Comparison with previous observational studies and caveats of our analysis}
\label{sect:dis:things2obs}

As discussed in \S\ref{sect:intro}, power spectra analysis of the cold gas content of galaxies is common in the literature. In most, if not all of these studies, the power spectra are well fit by single or broken power-laws from scales of a few 100 pc to kpc scales, in agreement with our analysis in \S\ref{sect:DPS}. The power law index for our sample of \things galaxies (NGC 628, 3521, 4736, 5055, 5457, 6946) match the large scale ($>$ few $100\pc$) results recovered from our simulation of a Milky Way like spiral galaxy when feedback is present, with $\alpha\sim 2-2.5$. Without feedback the spectra are too shallow to be compatible with observations.

We note that for four of the galaxies (NGC 628, 3521, 4736 and 5055) we find a larger value of $\alpha$ than those reported by \cite{Dutta:2008aa, Dutta:2009ab, Dutta:2013aa} and \cite{Dutta:2013ab} but a reasonable match to the indices quoted in \cite{Walker:2014aa}. A similar difference in power law index was found for our SMC model, which featured an almost single power-law of with $\alpha=1.8$, compared to the derived $\alpha=2.85$ from observations \cite{Stanimirovic:1999aa}. In contrast to our findings, \HI{} power spectra in local dwarf irregular galaxies in the Little \things sample are found to be better fit with steeper power laws \citep[][]{Zhang:2012aa} compared to spiral galaxies \citep[][]{Dutta:2013aa}. 

While it is tempting to discuss this, and other trends, further, we refrain from doing so as the value of the power law index is known to depend on the adopted definition of the power spectrum, velocity channel width \citep[][]{LazarianPogosyan2000}, integrated intensity maps vs. single-velocity-channel maps \citep[][]{Padoan2006} etc., which differ significantly in the literature. For example, we have confirmed that our definition on $\Pk$ agrees with \cite{Dutta:2008aa, Dutta:2009ab, Dutta:2013aa} and \cite{Dutta:2013ab} (private communication), but find a difference in method; in this work we have made use the zeroth moment maps available from the \things data archive, while \cite{Dutta:2008aa, Dutta:2009ab, Dutta:2013aa, Dutta:2013ab} calculates the \emph{visibility}, defined as the Fourier transform of the sky brightness and then the power spectrum from this quantity. Recent work by Nandakumar \& Dutta (in prep) has demonstrated that using images from radio interferometry produces systematically \textit{larger} values of $\alpha$ than a visibility based method. They conclude that this difference is due to a noise bias that cannot easily be separated from the images and therefore any power spectrum from such a map would contain power from both the galaxy and noise. 

To allow for a meaningful comparison between simulations and observations it is important to analyse both in a similar way as possible, as was attempted in this work, but a complete homogenisation of literature results is beyond the scope of this paper.

%-------------------------------------------------------------------------------------------------------------------------------------------------------------------------------------------
%--------------------------------------------------------------------------- Section: Conclusion ------------------------------------------------------------------------------------

\section{Conclusion}
\label{sect:con}
In this work we study the role of stellar feedback in shaping the density and velocity structure of neutral hydrogen  (\HI) in disc galaxies. To achieve this, we use \things data to compute \HI{} density and kinetic energy power spectra for 6 local spiral galaxies, and compare these to high resolution ($\sim 4.6\pc$) hydro+$N$-body simulations, using the AMR code \ramses, of entire disc galaxies. We carry out simulations of Milky Way, LMC and SMC-like galaxies, with and without stellar feedback, in order to identify differences in the interstellar medium. Our key results are summarised below.

\begin{itemize}

	\item Combined with gravity and shear, stellar feedback shapes the observed density field of galaxies, as illustrated through power spectra of \HI{} gas. Feedback creates a steepening of the power spectra on spatial scales below $\sim 1-2\kpc$, with $\alpha\sim 2.5$, in agreement with local spirals from \thingstext. This match is achieved by feedback preventing regions of very high densities ($\rho>10^4$cm$^{-3}$) from dominating the density field, and instead allowing for star formation in gas of average densities $\rho\sim 100\cc$, typical of observed GMCs of sizes $\sim 10-100\pc$.

	\item	The large scale shape of the \HI{} power spectra ($\gtrsim$ few $1\kpc$) is insensitive to stellar feedback and is controlled by the large scale distribution of gas, i.e. the extent of \HI{} at large galactic radii. 
	
	\item Line-of-sight \HI{} kinetic energy power spectra ($E(k)$) from simulation with feedback are in good agreement with observations up to kpc-scales for a majority of the studied spiral galaxies. Simulations without feedback  under-predict the observed kinetic energy present on large scales ($\gtrsim 0.5\kpc$), with excessive small scale power due the presence of dense star forming clouds. 	
	
	\item The inclination of a galaxy can have a significant impact on the measured line-of-sight \HI{} kinetic energy power spectra of a galaxy, as the energy in galactic rotation dominates over turbulent energy. Correcting for this is crucial in order to use $E(k)$ to differentiate different feedback models. By contrast, the density power spectra is only weakly affected for the moderate inclination angles ($i\lesssim 40\degree$) investigated in this work.

	\item In 3D, simulations with feedback produce kinetic energy spectra $E(k)\propto k^{-2}$, as expected for super-sonic turbulence, on scales $\lesssim$ few $100\pc$, with a break at large scales possibly related to disc thickness. This can only be achieved if feedback acts as a mechanism for moving gas from the small to large scales, where is it then free to collapse down to small scales again. Without a mechanism such as feedback to redistribute gas, it accumulates at small to medium scales ($\ell\lesssim 300 \pc$).
	
	\item Without feedback, the ISM roughly reaches equipartition in terms of the fraction of kinetic energy in solenoidal motions ($2/3$) vs. compressive ($1/3$). With stellar feedback, the fraction of energy in compressive modes increases to $\gtrsim 45\%$ for the Milky Way model, with a similar trend, but weaker effect, in the  SMC and LMC models.
		
\end{itemize}

To conclude, on top of gravity and shear, stellar feedback is a major driver of the density and energy structure of the ISM up to kpc scales, and these effects can be quantified using density and energy power spectra of \HI{} gas. 

Finally, despite omitted physics, such as magnetohydrodynamics, cosmological context and self-consistent modelling of \HHH, we find a good agreement with observations, suggesting these are next-to-leading order effects. We will revisit these areas in future work.

%-------------------------------------------------------------------------------------------------------------------------------------------------------------------------------------------
%--------------------------------------------------------------------------- Section: Acknowledgements -------------------------------------------------------------------------

\section*{acknowledgments}
We thank the anonymous referee for valuable comments. This work made use of `\thingstext', `The \HI{} Nearby Galaxy Survey' \citep{THINGSpaper}. 
We thank Denis Erkal for his advice on using \things and Prasun Dutta for fruitful discussion on the definition of power spectra. 
KG would like to thank the University of Surrey for his studentship. 
O.A. and J.R. would like to acknowledge support from STFC consolidated grant ST/ M000990/1 and O.A. acknowledges support from the Swedish Research Council (grant 2014- 5791). 
FR acknowledges support from the European Research Council through grant ERC-StG-335936.

%-------------------------------------------------------------------------------------------------------------------------------------------------------------------------------------------
%--------------------------------------------------------------------------- Bibliography ----------------------------------------------------------------------------------------------

\bibliographystyle{mn2e}
\bibliography{ref0}

%%-------------------------------------------------------------------------------------------------------------------------------------------------------------------------------------------
%%-------------------------------------------------------------------------------------------------------------------------------------------------------------------------------------------

\end{document}